\documentclass[%
 reprint,
 amsmath,amssymb,
 aps,
]{revtex4-2}

\usepackage{graphicx}% Include figure files
\usepackage{dcolumn}% Align table columns on decimal point
\usepackage{bm}% bold math
\usepackage{hyperref}% add hypertext capabilities
\usepackage[mathlines]{lineno}% Enable numbering of text and display math
\usepackage[thickspace,amssymb]{SIunits}
\usepackage{xcolor}
\usepackage{multirow}
\usepackage{url}
\usepackage{verbatim,upgreek} % added for insert comments and not italik greek letters
\usepackage{lipsum}

\begin{document}

\preprint{Physical Review C}

%\title[]{Double Differential Fragmentation Cross Section at Large Angles of Carbon ion beams on Composite Targets of interest for Particle Therapy}% Force line breaks with \\
\title{Cross Section Measurements of Large Angle Fragments Production in the Interaction of Carbon Ion Beams with Thin Targets}
%\thanks{Footnote to title of article.}
%\author{Y. Dong and I. Mattei on behalf of the FOOT Collaboration}
%\affiliation{INFN Section of Milan, Milan, Italy}%Lines break automatically or can be forced with \\

\author{Y. Dong$^{1}$, I. Mattei$^{1}$, A. Alexandrov$^{2}$, B. Alpat$^{3}$, G. Ambrosi$^{3}$, S. Argir\`o$^{4,5}$, M. Barbanera$^{3}$, N. Bartosik$^{5}$, G. Battistoni$^{1}$, M.G. Bisogni$^{6,7}$, V. Boccia$^{2,8}$, F. Cavanna$^{5}$, P. Cerello$^{5}$, E. Ciarrocchi$^{6,7}$, A. De Gregorio$^{9,10}$, G. De Lellis$^{2,8}$, A. Di Crescenzo$^{2,8}$, B. Di Ruzza$^{11,12}$,  M. Dondi$^{13,14}$, M. Donetti$^{5,15}$, M. Durante$^{8,16}$, R. Faccini$^{9,10}$, V. Ferrero$^{5}$, C. Finck$^{17}$, E. Fiorina$^{5}$, M. Francesconi$^{2}$, M. Franchini$^{13,14}$, G. Franciosini$^{10,18}$, G. Galati$^{19,12}$, L. Galli$^{7}$, M. Ionica$^{3}$, A. Iuliano$^{2,8}$, K. Kanxheri$^{3,20}$, A.C. Kraan$^{7}$, C. La Tessa$^{21,22}$, A. Lauria$^{2,8}$, E. Lopez Torres$^{5,23}$, M. Magi$^{10,18}$, A. Manna$^{13,14}$, M. Marafini$^{10,24}$, M. Massa$^{7}$, C. Massimi$^{13,14}$,  A. Mengarelli$^{13}$, A. Mereghetti$^{15}$, T. Minniti$^{25,26}$, A. Moggi$^{7}$, M.C. Montesi$^{2,27}$, M.C. Morone$^{25,26}$, M. Morrocchi$^{6,7}$, S. Muraro$^{1}$, N. Pastrone$^{5}$, V. Patera$^{10,18}$, F. Peverini$^{3,20}$, F. Pennazio$^{5}$, C. Pisanti$^{13,14}$, P. Placidi$^{3,28}$, M. Pullia$^{15}$, L. Ramello$^{5,29}$, C. Reidel$^{16}$, R. Ridolfi$^{13,14}$, L. Sabatini$^{30}$, L. Salvi$^{3,20}$, C. Sanelli$^{30}$, O. Sato$^{31}$, S. Savazzi$^{15}$, L. Scavarda$^{31}$, A. Schiavi$^{10,17}$, C. Schuy$^{15}$, E. Scifoni$^{21}$, L. Servoli$^{3}$, G. Silvestre$^{3}$, M. Sitta$^{5,32}$, R. Spighi$^{12}$, E. Spiriti$^{29}$, L. Testa$^{9,10}$, V. Tioukov$^{2}$, S. Tomassini$^{30}$, F. Tommasino$^{21,34}$, M. Toppi$^{10,18,*}$, A. Trigilio$^{30}$, G. Traini$^{10}$, G. Ubaldi$^{13,14}$, A. Valetti$^{4,5}$, M. Vanstalle$^{17}$, M. Villa$^{13,14}$, U. Weber$^{16}$, R. Zarrella$^{13,14}$, A. Zoccoli$^{13,14}$}
\author{A. Sarti$^{10,18}$}

\affiliation{$^{1}$INFN Section of Milano, Milano, Italy}
\affiliation{$^{2}$INFN Section of Napoli, Napoli, Italy}
\affiliation{$^{3}$INFN Section of Perugia, Perugia, Italy}
\affiliation{$^{4}$University of Torino, Department of Physics, Torino, Italy}
\affiliation{$^{5}$INFN Section of Torino, Torino, Italy}
\affiliation{$^{6}$University of Pisa, Department of Physics, Pisa, Italy}
\affiliation{$^{7}$INFN Section of Pisa, Pisa, Italy}
\affiliation{$^{8}$University of Napoli, Department of Physics "E.~Pancini", Napoli, Italy}
\affiliation{$^{9}$University of Rome La Sapienza, Department of Physics, Rome, Italy}
\affiliation{$^{10}$INFN Section of Roma 1, Rome, Italy}
\affiliation{$^{11}$University of Foggia, Foggia, Italy}
\affiliation{$^{12}$INFN Section of Bari, Bari, Italy}

\affiliation{$^{13}$INFN Section of Bologna, Bologna, Italy}
\affiliation{$^{14}$University of Bologna, Department of Physics and Astronomy, Bologna, Italy}
\affiliation{$^{15}$CNAO Centro Nazionale di Adroterapia Oncologica, Pavia, Italy}
\affiliation{$^{16}$Biophysics Department, GSI Helmholtzzentrum f\"ur Schwerionenforschung, Darmstadt, Germany}
\affiliation{$^{17}$Universit\'e de Strasbourg, CNRS, IPHC UMR 7871, F-67000 Strasbourg, France}
\affiliation{$^{18}$University of Rome La Sapienza, Department of Scienze di Base e Applicate per l'Ingegneria (SBAI), Rome, Italy}
\affiliation{$^{19}$University of Bari, Department of Physics, Bari Italy}

\affiliation{$^{20}$University of Perugia, Department of Physics and Geology, Perugia, Italy}
\affiliation{$^{21}$University of Miami, Radiation Oncology, Miami, FL, United States}
\affiliation{$^{22}$Trento Institute for Fundamental Physics and Applications, Istituto Nazionale di Fisica Nucleare (TIFPA-INFN), Trento, Italy}
\affiliation{$^{23}$CEADEN, Centro de Aplicaciones Tecnologicas y Desarrollo Nuclear, Havana, Cuba}
\affiliation{$^{24}$Museo Storico della Fisica e Centro Studi e Ricerche Enrico Fermi, Rome, Italy}
\affiliation{$^{25}$University of Rome Tor Vergata, Department of Physics, Rome, Italy}
\affiliation{$^{26}$INFN Section of Roma Tor Vergata, Rome, Italy}
\affiliation{$^{27}$University of Napoli, Department of Chemistry, Napoli, Italy}
\affiliation{$^{28}$University of Perugia, Department of Engineering, Perugia, Italy}
\affiliation{$^{29}$University of Piemonte Orientale, Department for Sustainable Development and Ecological Transition, Vercelli, Italy}
\affiliation{$^{30}$INFN Laboratori Nazionali di Frascati, Frascati, Italy}
\affiliation{$^{31}$Nagoya University, Department of Physics, Nagoya, Japan}
\affiliation{$^{32}$ALTEC, Aerospace Logistic Technology Engineering Company, Corso Marche 79, 10146 Torino, Italy}
\affiliation{$^{33}$University of Piemonte Orientale, Department of Science and Technological Innovation, Alessandria, Italy}
\affiliation{$^{34}$University of Trento, Department of Physics, Trento, Italy, \\
$^{*}$Corresponding Author: Marco Toppi, marco.toppi@uniroma1.it}
%\affiliation{$^{*}$Corresponding Author: Marco Toppi, marco.toppi@uniroma1.it}
\collaboration{The FOOT Collaboration}%\noaffiliation

%%%%%%%%%

%\author{}%
%\email{Second.Author@institution.edu.}
%\affil[iation{ 
%Authors' institution and/or address%\\This line break forced with \textbackslash\textbackslash
%}%

% \homepage{http://www.Second.institution.edu/~Charlie.Author.}
%\affil[iation{%
%Second institution and/or address%\\This line break forced% with \\
%}%


\begin{abstract}
The fragmentation cross sections of carbon ion beams with kinetic energies of $115 - 353~\text{MeV/u}$ impinging on thin targets of graphite (C), polyvinyl-toluene (C$_9$H$_{10}$) and PMMA (C$_2$O$_5$H$_8$) have been measured at 90$^{\text{o}}$ and 60$^{\text{o}}$ at the CNAO particle therapy center (Pavia, Italy). 
The presented measurements are a complete reanalysis by the FOOT collaboration of already published elemental cross section on composite targets, in order to refine the analysis, improve the systematic uncertainties and show the comparison with the FLUKA Monte Carlo code calculations. In this work, the kinetic energy at production of measured fragments has been completely redefined, together with the efficiencies computation. The new analysis strategy has been successfully validated against the Monte Carlo cross sections. Two detection arms were positioned at two different angles to perform the measurement at 90$^{\text{o}}$ and 60$^{\text{o}}$. The fragment species have been identified in charge (Z$_{id}$ = H) and mass (M$_{id}$ = $^1$H, $^2$H, $^3$H) combining the information of the deposited energy in thin plastic scintillators, of the deposited energy in a thick LYSO crystal and of the fragments Time of Flight (ToF) measurement. The ToF was also used to compute the fragments measured kinetic energy. The cross sections are presented as a function of the fragments kinetic energy at production thanks to an unfolding technique applied to the data.
\end{abstract}

%Use showkeys class option if keyword
\maketitle

\section*{\label{sec:intro}Introduction}

Particle Therapy (PT) is the external radiation therapy technique that exploits protons and carbon ion beams to treat especially deep-seated solid tumors close to organs at risk~\cite{paper1}. In particular, carbon ions are used to treat radio-resistant tumors thanks to their higher biological effectiveness in killing cancerous cells with respect to photons and protons~\cite{paper2}, but hadrons with a mass number A~$>$~1 may undergo fragmentation in the nuclear interaction with patients tissue nuclei. Fragments with high mass will be produced at small angles with respect to the projectile incident direction, causing the dose tail beyond the Bragg peak, while low mass fragments such $^1$H, $^2$H and $^3$H can be produced even at large angles, depositing their dose far from the beam trajectory. The knowledge of the production of Z~=~1 fragments at large angles is also of interest for beam range monitoring techniques based on the detection of charged secondary fragments~\cite{paperDP} and for radio-protection purposes on long term space missions~\cite{paperSpace1,paperSpace2}. Treatment plans based on simulations both with analytical and Monte Carlo (MC) approaches~\cite{papersMC1} suffer for the uncertainties on the Relative Biological Effectiveness (RBE) assessment, due to the large uncertainties on the knowledge of the production fragmentation cross section of 80~-~400~MeV/u $^{12}$C ion beam, both at experimental level and in the related calculation models\cite{muraro2020}. Several measurements %up to 95~MeV/u beam energy, 
with a detection angle below 45$\degree$,  and only a small number of beam-target-energy combinations, have been performed by other research groups~\cite{Webber90,Dodouet2013,Dodouet201406,Zeitlin2011,Pleskac2012,Toppi2016}.

In this paper, the cross section measurements for the production of Z~=~1 fragments (protons, deuterons and tritons), detected at large angles (90$\degree$ and 60$\degree$), from the interaction of $115 - 353~\text{\rm MeV/u}$ $^{12}$C ion beams with a graphite (C), polyvinyl-toluene (EJ-212 from Scionix~\cite{scionixref}, CH in the following) and polymethyl-methacrylate (PMMA) targets are reported. The experimental data of the cross section of $^{12}$C ion beam on carbon (C), oxygen (O) and hydrogen (H) elemental targets at large angle have been already published~\cite{paperfoot} by the FOOT collaboration~\cite{footexp}, exploiting the composite targets subtraction method from the cross section of $^{12}$C ion beam impinging on C, CH and PMMA targets, from a data taking performed at the CNAO therapy center (Pavia, Italy). In this work, the same dataset is reanalyzed, driven by advancements in the whole analysis strategy. In particular, improvements have been made in the efficiency computation, in the extraction of the kinetic energy through a new unfolding technique and in the systematic uncertainties evaluation. In Section~\ref{sec:1} the experimental setup and configurations are described, in Section~\ref{sec:method} the data analysis strategy is presented. %~\footnote{Fondazione CNAO - Centro Nazionale di Adroterapia Oncologica, Pavia, Italy}. 
The computation of the fragment kinetic energy at production has been implemented exploiting an unfolding technique of the measured fragment kinetic energy (see Section~\ref{sec:unfmat}), while in the already published data an analytic function was applied for the measured kinetic energy correction. Moreover, instead of computing a fragment identification efficiency averaged on the kinetic energy at production of fragments, in the presented analysis the fragment identification efficiency is modulated as a function of the fragment production kinetic energy (see Section~\ref{sec:effimix}). The systematic uncertainty evaluation is discussed in Section~\ref{sec:syserr}. In Section~\ref{sec:result} the results are reported and the comparison with FLUKA Monte Carlo code~\cite{fluka1,fluka2} predictions is also shown for the first time and discussed in Section~\ref{sec:discuss}.

\section{\label{sec:1}Experimental Configurations}

The double differential fragmentation cross sections of $^{12}$C ion beam over low density graphite sheet (C), polyvinyl-toluene (polyvinyl in Table~\ref{tab:target} - CH) and PMMA thin targets (see Table~\ref{tab:target}) have been measured, exploiting five beam energies: 115~MeV/u, 150~MeV/u, 221~MeV/u, 279~MeV/u and 351~MeV/u. The beam intensity was the therapeutical one ($\sim$~10$^8$~ions/s). Each target, with atomic mass number $A_Y$, was placed at 45$\degree$ with respect to the incoming beam direction ($th_{Y} = Thickness \cdot \sqrt{2}$), and was impinged by $\sim$~5~$\cdot$~10$^{10}$ ions.
\begin{table}[ht]
\begin{center}
\begin{tabular}{|l|c|c|c|c|c|}
\hline
Target  &  Composition &  Thickness & $th_Y$  & Density & $A_Y$\\
  &   &  [$mm$] &  [$mm$] & [$g/cm^3$] & [$u$]\\
\hline
\hline
Graphite & C  & $1$ & $1.4$ & $0.94$ & 12.01\\
Polyvinyl & C$_9$H$_{10}$  & $2$&   $2.8$ & $1.024$ & 118.18\\
PMMA & C$_5$O$_2$H$_8$ & $2$ & $2.8$ & $1.19$ & 100.12\\
\hline
\end{tabular}
\end{center}
\caption{Targets composition and parameters~\cite{paperfoot}.} 
\label{tab:target}
\end{table}

\begin{figure}[h!]
    \centering
    \includegraphics[width=1.\linewidth]{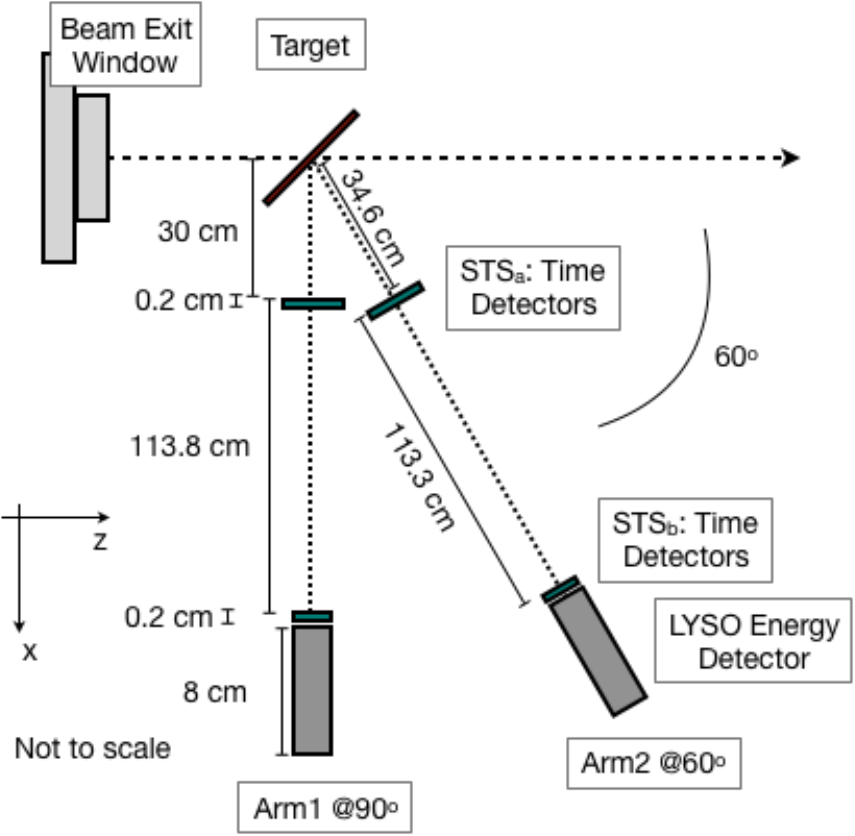}
    \caption{Sketch of the experimental setup (not to scale)~\cite{paperfoot}.}
    \label{fig:setup}
\end{figure}

The experimental setup is shown in Fig.~\ref{fig:setup}: 
at 90$\degree$ and 60$\degree$, two detection arms (Arm1 and Arm2) were placed for the simultaneous detection of secondary charged fragments. In each arm, two thin plastic scintillators, 2~mm thick ($STS_a$ and $STS_b$), were used for the Time of Flight (ToF) and energy loss measurement of the fragments, while a LYSO crystal 4~$\times$~4~$\times$~8~cm$^3$ was used as calorimeter for the fragments kinetic energy measurements. The Data AcQuisition (DAQ) trigger was the logic OR of the trigger of Arm1 and Arm2. The trigger of Arm1(2) was the logic AND of the STS$_a$ and STS$_b$ discriminated signals of Arm1(2) (more details on the experimental setup are given in~\cite{paperfoot}). For each trigger, the fragments considered for the analysis are the ones whose correspond to an energy release in both the STSs and in the LYSO.

\section{Methods}
\label{sec:method}
The double differential cross section computed as a function of the fragment kinetic energy at production ($E_{k}$) and measured at $\theta = 60\degree, 90\degree$ is defined as:

\begin{equation}
\frac{d^2\sigma}{d\Omega dE_{k}}(^A _ZX)= \frac{N_{^A _ZX}(E_{k})}{N_{Y}~\Delta\Omega~\Delta E_{k}~N_{^{12}C}~\epsilon(E_k)}~.
    \label{eq:XS}
\end{equation}

$N_{^A _ZX} (E_{k})$ is the number of fragments with a specific atomic number $Z$ and mass number $A$, in each kinetic energy bin $E_{k}$; $N_{Y}$ is the number of scattering centers per unit surface; $\Delta \Omega$ is the solid angle of the fragments at production seen and reconstructed by the LYSO detector; $\Delta E_{k}$ is the fragment kinetic energy bin size; $N_{^{12}C}$ is the number of incoming carbon ions and $\epsilon(E_k)$ is the total efficiency.

The number of scattering centres in a $Y$ target per unit surface is defined as:
\begin{equation}
N_{Y} = \frac{\rho_Y \cdot th_Y \cdot N_A}{A_Y}
\end{equation}\\
where $A_Y$ is the atomic mass number, $N_A$ the Avogadro number, $\rho_Y$ the target density and $th_Y$ is the thickness of the crossed target (see Table~\ref{tab:target}). 
$\Delta \Omega$ has been computed by means of the MC simulation, taking into account the spatial distribution of the beam and the multiple scattering underwent by fragments before reaching the LYSO detector. %The obtained $\Delta \Omega$ is $\Delta \Omega_{90} = 19$~msr and $\Delta \Omega_{60} = 17$~msr, respectively for the angular setup of Arm1 and Arm2. 
The number of carbon ions $N_{^{12}C}$ is provided by the CNAO dose delivery system from the released charged measured by the ionization chambers. A $4\%$ relative uncertainty on $N_{^{12}C}$ is related to the current measurement precision and to the dose-current conversion systematic uncertainty~\cite{Donetti1,Donetti2,Donetti3} and it has been accounted in the final result as a systematic uncertainty. 

The total efficiency $\epsilon(E_k)$ is factorized in three terms, each one depending on the production kinetic energy: 
\begin{equation}
\epsilon(E_k) = \epsilon_{Rec} \cdot \epsilon_{PID} \cdot \epsilon_{DT}
\label{eq:effi}
\end{equation}
where $\epsilon_{Rec}$ includes the geometrical, trigger and detection efficiency of Z~=~1 fragment (see sec.~\ref{sec:effidet}), $\epsilon_{PID}$ is the particle identification efficiency (see sec.~\ref{sec:effimix}) and $\epsilon_{DT}$ is the dead time efficiency, in order to take into account for the DAQ dead time, which depends on the beam rate and was measured to be in the 2 - 8$\%$ range, as reported in~\cite{paperfoot}. %more on epsDT! say that is the mean of the distribution and therefore is corrected as <1/epsDT>.
To evaluate $\epsilon_{Rec}$ and $\epsilon_{PID}$ and other corrections defined in the following sections, FLUKA MC simulations of the geometrical setup have been performed, one for each target type and beam energy.

\subsection{Fragment Identification}
\label{sec:fid}
The number of specific fragments, $N_{^A _ZX}$ as a function of $E_{k}$, \textit{i.e.} the fragment kinetic energy at production, is evaluated following the equation:
\begin{equation}
N_{^A _ZX} (E_{k}) = \boldsymbol{U}\cdot(N_{^A _ZX} (E^{m}_{k}) \cdot p(E^{m}_{k}))
\label{eq:nfrag}
\end{equation}

where $\boldsymbol{U}$ is the unfolding matrix (see sec.~\ref{sec:unfmat}), $p(E^{m}_{k})$ is the \textit{purity} (see eq.~\ref{eq:pure}), $N_{^A _ZX} (E^{m}_{k})$ is the number of ${^A _ZX}$ fragment as a function of the measured kinetic energy:
$$
E_{k}^{m} = m_i c^2\cdot (\gamma_i - 1);
$$
with 
$$
\gamma_i = (1 - \beta_i^2)^{-1/2}\quad ,\quad \beta_i = L/(ToF_i \cdot c).
$$
$m_i$ is the mass of the fragment $i=p (^1H), d (^2H), t (^3H) $, $L$ is the distance between STS$_a$ and STS$_b$, $c$ is the speed of light and $ToF_i$ is the measured time of flight of the $i$ fragment.

The number of ${^A _ZX}$ fragment has been evaluated after the particle identification ($PID$) in charge Z and mass A, following the same procedure described in~\cite{paperfoot}: to identify the fragment charge (Z=1), the information on the energy released in the STSs, by means of a charge-to-digital converter (QDC) module (CAEN V792), has been combined with the LYSO QDC and ToF measurement; the fragment mass (A=1,2,3) has been identified combining the LYSO QDC with the measurement of 1/ToF. 
Once $N_{^A _ZX} (E^{m}_{k})$ has been evaluated, it has been corrected for the \textit{purity}. As described in~\cite{paperfoot}, the same PID algorithm has been implemented in the MC simulation using the Energy and ToF experimental resolution to tune the MC response to data. \textit{Purity} has been computed thanks to the FLUKA Monte Carlo simulation of the full geometrical setup for each target-beam configuration and it is defined as:
\begin{equation}
\label{eq:pure}
p(E^{m}_{k}) = \frac{N^{MC}_{^A _ZX} (E^{m}_{k})_{PID}}{N^{recoMC}_{^A _ZX} (E^{m}_{k})_{PID}}.
\end{equation}
$E^m_k$ is the kinetic energy obtained by ToF, as in experimental data, $N^{MC}_{^A _ZX} (E^{m}_{k})_{PID}$ is the number of correctly identified $^A _ZX$ fragments and $N^{recoMC}_{^A _ZX} (E^{m}_{k})_{PID}$ is the number of reconstructed $^A _ZX$ fragments after identification. \textit{Purity} values range between $95~-~100\%$. Clearly, the discrepancies between data and MC response are a source of systematic effects which enter both in PID and purity corrections. These contributions have been included in the overall systematic uncertainty and will be discussed in detail in sec.~\ref{sec:syserr}.

\subsubsection{Unfolding of measured kinetic energy}
\label{sec:unfmat}

\begin{figure}[h!]
    \centering
    \includegraphics[width=1.\linewidth]{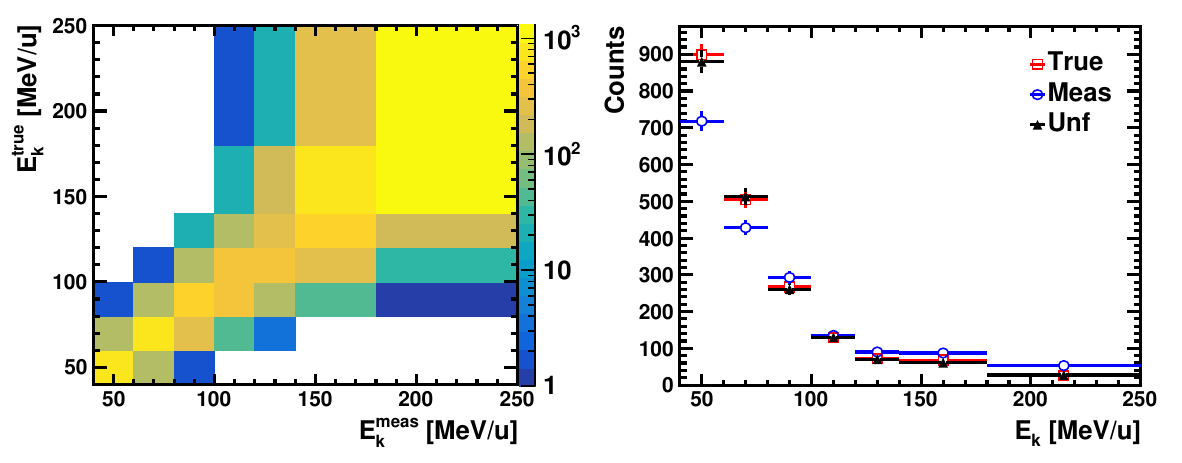}
    \caption{Left: unfolding matrix $\boldsymbol{U}$ of $E_k^{}$ as a function of $E_k^{meas}$, obtained from a full MC simulation, for the case of protons detected at 90$\degree$ and produced by a 351~MeV/u $^{12}$C ion beam impinging on the PMMA target. Right: tuning of the unfolding procedure. The unfolding matrix (left) is applied to the MC reconstructed protons distribution (blue circles) in order to obtain the unfolded distribution (black triangles). The unfolded kinetic energy at production has to be compared to the  MC protons kinetic energy at production (red squares).}
    \label{fig:unf}
\end{figure}

After the \textit{purity} correction, to obtain the number of ${^A _ZX}$ fragment as a function of the kinetic energy of the fragment at production ($N_{^A _ZX} (E_{k})$ in eq.~\ref{eq:nfrag}), an unfolding technique, based on the \textit{RooUnfoldBayes} method of the RooUnfold package~\cite{roounf} based on the ROOT software~\cite{rootcern}, has been applied to the measured kinetic energy spectrum, due to the fact that a fragment loses energy before exiting the target and in air before being detected. The unfolding matrix $\boldsymbol{U}$ has been computed from 
%\textcolor{red}{a MC simulation, where protons, deuterons and tritons have been produced, separately, within the target, with a 4$\pi$ angular distribution and a uniform (\textit{FLAT}) kinetic energy spectrum between 5~MeV/u~-~1~GeV/u}. 
a MC simulation of the geometrical setup reproducing each experimental configuration of beam energy-target type. 
The unfolding matrix, gives the probability that a fragment with a measured kinetic energy had a certain production energy. 
The $\boldsymbol{U}$ matrix is then applied to the measured fragment yields ${^A _ZX}$, reconstructed in bin of kinetic energy, to obtain the fragment yields ${^A _ZX}$ in bin of production kinetic energy. As an example of the result, in Fig.~\ref{fig:unf} left, it is shown the matrix for the case of protons produced in PMMA and detected by Arm2 (60$\degree$). In Fig.~\ref{fig:unf} right, the unfolding matrix is applied to the Monte Carlo measured kinetic energy spectrum of reconstructed protons (blue circles), obtained from the MC simulation of the setup used also to compute the \textit{purity}. The unfolded energy spectrum (black triangles) is therefore obtained and compared to the  kinetic energy spectrum at production (red squares). The unfolding parameter of the \textit{RooUnfoldBayes} method, which is the number of the unfolding iteration \textit{niter}, is tuned exploiting the MC distributions as described, minimizing the $\chi^2$ between the  and the unfolded MC distributions. %A satisfactory convergence has been reached for \textit{niter}~=~2. 

%To evaluate the systematic uncertainty on the final result due to the unfolding method, the \textit{RooUnfoldIDS} method has been applied. The systematics due to this source of uncertainty is negligible with respect to other systematics (see sec.~\ref{sec:result}).
 
%\begin{figure}[h!]
%    \centering
%    \includegraphics[width=1.\linewidth]{FigUnf_FLAT.pdf}
%    \caption{Left: unfolding matrix $\boldsymbol{U}$ of $E_k^{}$ as a function of $E_k^{meas}$, obtained from a \textit{FLAT} MC simulation, for the case of protons produced in PMMA detected at 60$\degree$. Right: tuning of the unfolding procedure. The unfolding matrix (left) is applied to the MC reconstructed protons $E_k^{meas}$ distribution (blue circles) in order to obtain the unfolded distribution $E_k^{unf}$ (black triangles), which is the unfolded kinetic energy at production, to be compared to the  MC protons kinetic energy at production $E_k^{}$ (red squares).}
%    \label{fig:unf}
%\end{figure}

\subsection{Efficiency Evaluation}
\subsubsection{Reconstruction Efficiency}
\label{sec:effidet}
The reconstruction efficiency is the convolution of geometrical, trigger and detection efficiencies. It has been computed as a function of the fragment kinetic energy at production, following the equation:
\begin{equation}
\epsilon_{Rec} (E_k) = \frac{N^{recoMC}_{_Z^AX}(E_k)_{TE}}{N^{trueMC}_{_Z^AX}(E_k)_{\Delta\Omega}}
\label{eq:effidet}
\end{equation}
where $N^{recoMC}_{_Z^AX}(E_k)_{TE}$ is the number of  $_Z^AX$ fragments MC reconstructed as experimental data after the trigger selection ($T$) and detectors energy thresholds cuts ($E$). No PID is used at this level, the fragment identification is done at  MC generator level. The trigger is defined in simulation in the same way as in experimental data acquisition, that is the time coincidence %(in a time window of 150~ns)
between the two STSs. The same is done for energy threshold cuts, implemented in the MC analysis as they are in experimental data. %: the STSs have an energy threshold set between 2-5~MeV, the LYSO energy threshold is set up to 24~MeV, to reduce neutral particle background and avoid signal from the lower energy secondary protons which are of no interest for range monitoring applications. 
$N^{trueMC}_{_Z^AX}(E_k)_{\Delta\Omega}$ is the number of generated $_Z^AX$ fragments produced in the $\Delta\Omega$ seen by the LYSO detector. The denominator of the reconstruction efficiency corresponds to the MC prediction of the yield of the produced fragment $_Z^AX$ to be compared with the measured experimental data (see sec~\ref{sec:result}). 

\begin{figure}[h!]
    \centering
    \includegraphics[width=.8\linewidth]{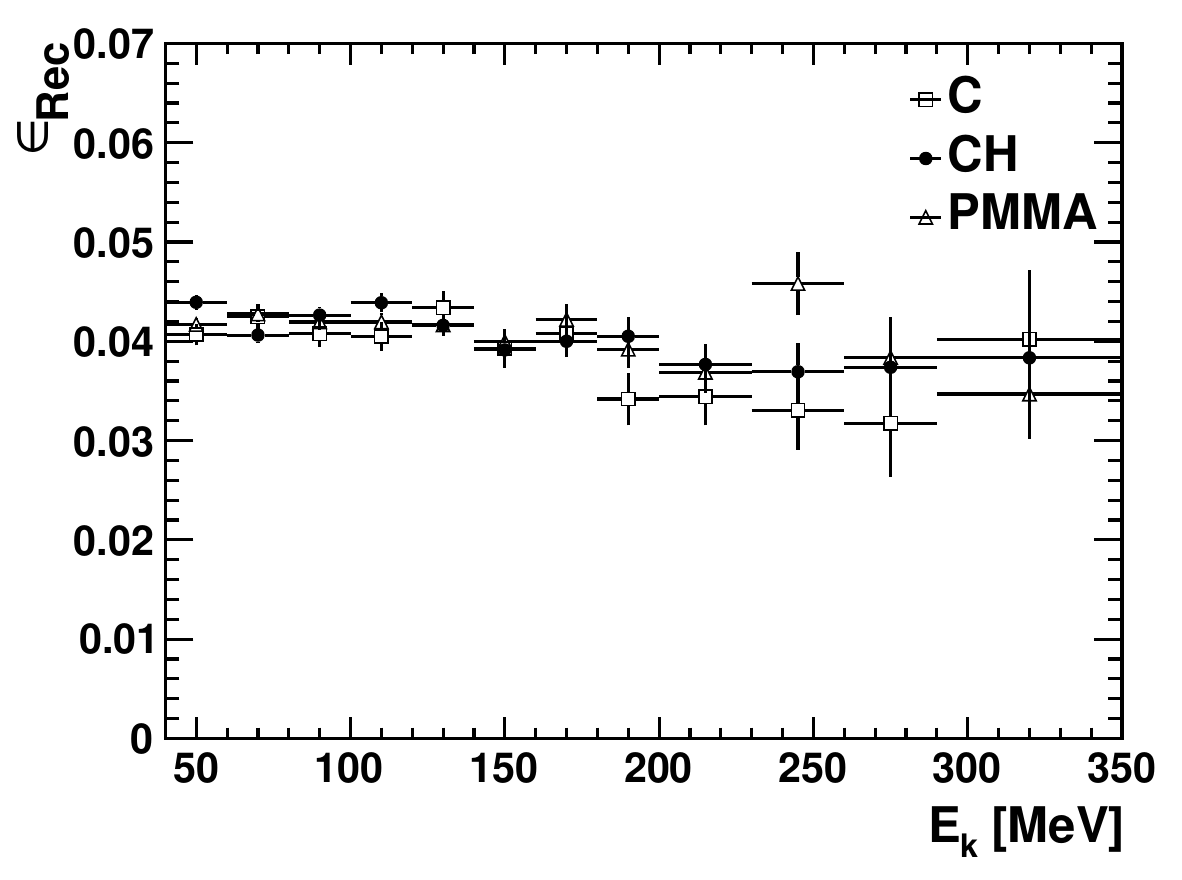}
    \caption{Reconstruction efficiency (convolution of geometrical, trigger and detection efficiencies) as a function of the kinetic energy at production of the detected proton fragments. The shown efficiency is computed from the FLUKA MC simulation of the 351~MeV/u $^{12}$C ion beam impinging on C, CH and PMMA targets, with fragments detected at 60$\degree$.}
    \label{fig:effidet}
\end{figure}

An example of the obtained $\epsilon_{Rec}(E_k)$ for secondary proton fragments detected at 60$\degree$ is shown in Fig.~\ref{fig:effidet}
for the case of the 351~MeV/u $^{12}$C ion beam impinging on C (open square), CH (full circle) and PMMA (open triangle) targets. No dependency of the efficiency to the target type is observed. The small absolute value is mostly due to the reduced angular acceptance of the setup.

\subsubsection{Particle Identification Efficiency}
\label{sec:effimix}
The particle identification efficiency ($\epsilon_{PID}$) has been computed as a function of the reconstructed $_Z^AX$ fragment kinetic energy at production, with the following equation:
\begin{equation}
\label{eq:effPID}
\epsilon_{PID} (E_k) =  \frac{N^{recoMC}_{_Z^AX}(E_k)_{PID}}{N^{recoMC}_{_Z^AX}(E_k)_{TE}}
\end{equation}
with $N^{recoMC}_{_Z^AX}(E_k)_{PID}$ is the number of MC fragments reconstructed in Z and A after the particle identification selection (see sec.~\ref{sec:fid}), implemented as it is in experimental data, while $N^{recoMC}_{_Z^AX}(E_k)_{TE}$ is the same as the numerator of $\epsilon_{Rec} (E_k)$ (see sec.~\ref{sec:effidet}). In Table~\ref{tab:effmix} the result of $\epsilon_{PID}$ for the identified proton fragments ($\epsilon_{PID}^{p}$) is reported for the case of MC simulation of 351~MeV/u $^{12}$C ion beam impinging on PMMA target and proton fragments detected at 60$\degree$.

%\begin{table}[ht]
%\begin{center}
%\begin{tabular}{| c || c | c | c |}
%\hline
%\rule[-4mm]{0mm}{1.cm}
% 60$\degree$ | $E_{kin}$ [MeV/u] & $\epsilon_{PID}^{p}$ [\%] & $\epsilon_{PID}^{d}$ [\%] & $\epsilon_{PID}^{t}$ [\%] \\
%\hline
%50 $\pm$ 10 & 98.4 $\pm$ 0.2 & 96 $\pm$ 1 & 94 $\pm$ 4 \\
%70 $\pm$ 10 & 96.7 $\pm$ 0.3 & 85 $\pm$ 3 & 100 $\pm$ 4 \\
%90 $\pm$ 10 & 94.7 $\pm$ 0.5 & 78 $\pm$ 6 & - \\
%110 $\pm$ 10 & 93.8 $\pm$ 0.5 & 47 $\pm$ 11 & - \\
%130 $\pm$ 10 & 91.7 $\pm$ 0.7 & 50 $\pm$ 17 & - \\
%150 $\pm$ 10 & 86 $\pm$ 1 & - & - \\
%170 $\pm$ 10 & 83 $\pm$ 1 & - & - \\
%190 $\pm$ 10 & 70 $\pm$ 2 & -& - \\
%215 $\pm$ 15 & 55 $\pm$ 3 &- & - \\
%245 $\pm$ 15 & 61 $\pm$ 4 & - & - \\
%275 $\pm$ 15 & 71 $\pm$ 5 & - & - \\
%320 $\pm$ 30 & 81 $\pm$ 6 & - & - \\
%\hline
%\end{tabular}
%\end{center}
%\caption{Particle identification efficiency $\epsilon_{PID}(E_k)$ as a function of the fragment production kinetic energy, for the case of MC simulation of 351~MeV/u $^{12}$C ion beam impinging on PMMA target and secondary fragments detected at 60$\degree$. The three columns show the $\epsilon^{i}_{PID}$, \textit{i.e.} the probability of identify the fragment $i$ in the $i-th$ fragment identification selection, with $i= p,d,t$.} 
%\label{tab:effmix}
%\end{table}

\begin{table}[ht]
\begin{center}
\begin{tabular}{| c || c | }
\hline
\rule[-4mm]{0mm}{1.cm}
 60$\degree$ | $E_{kin}$ [MeV/u] & $\epsilon_{PID}^{p}$ [\%] \\
\hline
50 $\pm$ 10 & 98.4 $\pm$ 0.2  \\
70 $\pm$ 10 & 96.7 $\pm$ 0.3  \\
90 $\pm$ 10 & 94.7 $\pm$ 0.5  \\
110 $\pm$ 10 & 93.8 $\pm$ 0.5 \\
130 $\pm$ 10 & 91.7 $\pm$ 0.7 \\
150 $\pm$ 10 & 86 $\pm$ 1 \\
170 $\pm$ 10 & 83 $\pm$ 1 \\
190 $\pm$ 10 & 70 $\pm$ 2  \\
215 $\pm$ 15 & 55 $\pm$ 3  \\
245 $\pm$ 15 & 61 $\pm$ 4 \\
275 $\pm$ 15 & 71 $\pm$ 5 \\
320 $\pm$ 30 & 81 $\pm$ 6 \\
\hline
\end{tabular}
\end{center}
\caption{Particle identification efficiency $\epsilon_{PID}(E_k)$ as a function of the fragment production kinetic energy, for the case of MC simulation of 351~MeV/u $^{12}$C ion beam impinging on PMMA target and secondary proton fragments detected at 60$\degree$. %The three columns show the $\epsilon^{i}_{PID}$, \textit{i.e.} the probability of identify the fragment $i$ in the $i-th$ fragment identification selection, with $i= p,d,t$.
} 
\label{tab:effmix}
\end{table}

The trend of $\epsilon_{PID}^{p}$ is due to the PID selection functions, that are done on the experimental data distribution of the deposited energy in LYSO (\textit{Eloss} in terms of LYSO QDC) as a function of 1/ToF, with ToF the fragments time of flight (see sec.~\ref{sec:fid}). When the kinetic energy of a particle exceeds a certain energy, the energy lost by that particle in the detector decreases as 1/ToF increases: the particle ``punches through'' the LYSO detector, meaning it is no longer fully contained within it. This effect is accurately reproduced by the Monte Carlo simulation, thanks to the MC distribution of the deposited energy in LYSO that has been tuned in order to reproduce the shape of the LYSO QDC of experimental data. This effect is taken into account by $\epsilon_{PID}$ (see Table~\ref{tab:effmix}), with the corresponding systematic uncertainty $sys_{PID}$ highlighting its impact (as described in sec.~\ref{sec:syserr}).

\subsection{Systematic uncertainty evaluation}
\label{sec:syserr}
A crucial aspect in the evaluation of the results (see sec.~\ref{sec:result}) is the assessment of the systematic uncertainty, that, in this analysis, is computed as a function of the fragment kinetic energy at production. 
The systematic uncertainty in the cross section measurement (as defined in eq.~\ref{eq:XS}) originates from the following main sources: %We point out that, as expected, the main contribution to $sys_{MC}$ is due to the unfolding procedure, \textit{i.e.} to the evaluation of the fragments kinetic energy at production from the measured kinetic energy. 

%We expect that one of the main contribution would come from the computation of the efficiencies, including the unfolding, which have been obtained by MC simulation. In order to assess the reliability in the determination of the applied efficiencies and unfolding, we performed a Monte Carlo consistency test. This is done by calculating the relative difference between the MC prediction computed with eq.~\ref{eq:XSmc} and the MC reconstructed differential cross section, computed following eq.~\ref{eq:XS}.

%\begin{equation}
%\frac{d^2\sigma^{recoMC}}{d\Omega dE_{k}}(^A _ZX)= \frac{N^{MC}_{^A _ZX}(E_{k})}{N_{Y}~\Delta E_{k}~\Delta \Omega~N^{MC}_{^{12}C}~\epsilon(E_k)}.
%    \label{eq:XSrecoMC}
%\end{equation}
%
%The number of reconstructed MC fragment is calculated as $N^{MC}_{^A _ZX}(E_{k}) = \boldsymbol{U}\cdot(N^{recoMC}_{^A _ZX} (E^{m}_{k}) \cdot p(E^{m}_{k}))$, following eq.~\ref{eq:nfrag}, with $N^{recoMC}_{^A _ZX} (E^{m}_{k})$ the number of reconstructed $^A _ZX$ fragment after the fragment identification. In eq.~\ref{eq:XSrecoMC}, $\epsilon(E_k)$ is the same of eq.~\ref{eq:effi}, except for $\epsilon_{DT}$ which accounts only for experimental data.
\begin{itemize}
\item[1.] Unfolding procedure ($sys_{unf}$)
\item[2.] $\Delta\Omega$ evaluation from MC simulation ($sys_{\Delta\Omega}$)
\item[3.] Particle identification selections ($sys_{PID}$)
\item[4.] Evaluation of the number of incoming ions $N_{^{12}C}$ (relative systematics 4$\%$ - see sec.~\ref{sec:method}). 
%\item[4.] Unfolding method: the use of \textit{RooUnfoldIDS} instead of \textit{RooUnfoldBayes} (relative systematics $<$1$\%$)
\end{itemize}

The dominant contribution comes from the unfolding procedure used to correct the migration of measured kinetic energy to production kinetic energy. This contribution has been evaluated within MC, even using a different unfolding technique (\textit{RooUnfoldIDS} instead of \textit{RooUnfoldBayes}) and its impact ranges between 0.1$\%$ to 13$\%$ for the most populated bins. This systematics in the previous published work~\cite{paperfoot} was completely underestimated, reminding that the energy correction in that work had been done with an analytic function.
$sys_{unf}$ is also the main source of uncertainty in the differential cross sections. 

The second contribution to the systematic uncertainty comes from the evaluation of $\Delta\Omega$, as described in sec.~\ref{sec:method}. $\Delta\Omega$ is computed by exploiting the MC simulation, taking into account the reconstructed fragments, since no tracking system is present in the experimental setup (see Fig.~\ref{fig:setup}). $\Delta\Omega$ affects the cross section normalization and the reconstruction efficiency evaluation, as shown in eq.~\ref{eq:XS} and eq.~\ref{eq:effidet}. Therefore, the impact of the $\Delta\Omega$ choice in the MC influences the final cross section results. For this reason, $\Delta\Omega$ has been varied from the default value and evaluated its impact on the systematic uncertainty $sys_{\Delta\Omega}$. This ranges from 1$\%$ up to 14$\%$ for the most populated bins. %, and it's comparable or below the main source of systematics $sys_{unf}$.  
%$\Delta\Omega$ is computed from the $\theta$ and $\phi$ polar angles distributions of fragments exiting the target and arriving at the LYSO detector, taking into account the distributions tails up to 4$\sigma$ from the nominal LYSO angular position ($\theta$ = 90$\degree$,60$\degree$; $\phi$ = 0$\degree$). Such solid angle is called $\Delta\Omega_4$ in the following. The systematic uncertainty on this uncertainty source has been computed by evaluating $\Delta\Omega$ considering the polar angles distributions tails up to 2$\sigma$ ($\Delta\Omega_2$), re-computing efficiencies and cross sections within the $\Delta\Omega_2$ selection. $sys_{\Delta\Omega}$ is, therefore, the relative difference between the cross section produced in $\Delta\Omega_4$ and $\Delta\Omega_2$. 

The systematic uncertainty on the differential cross section due to the PID selection ($sys_{PID}$) has been evaluated by varying of 1$\%$ the PID selection functions. The relative error due to this systematics ranges between 0.1$\%$ up to 6$\%$, depending on the fragment kinetic energy bin. The PID systematics has an impact in the fragment yields (see eq.~\ref{eq:XS} and sec.~\ref{sec:fid}), in the \textit{purity} (see eq.~\ref{eq:pure}) and in the PID efficiency evaluation (see eq.~\ref{eq:effPID}). The higher the fragment kinetic energy at production, the higher the relative systematics due to this source of uncertainty, as expected due to the bigger discrepancies between the PID selection bands and the \textit{Eloss} vs 1/ToF distributions at high energy. 

\section{Results}
\label{sec:result}
\begin{figure*}
\begin{center}
\includegraphics[width=0.8\linewidth]{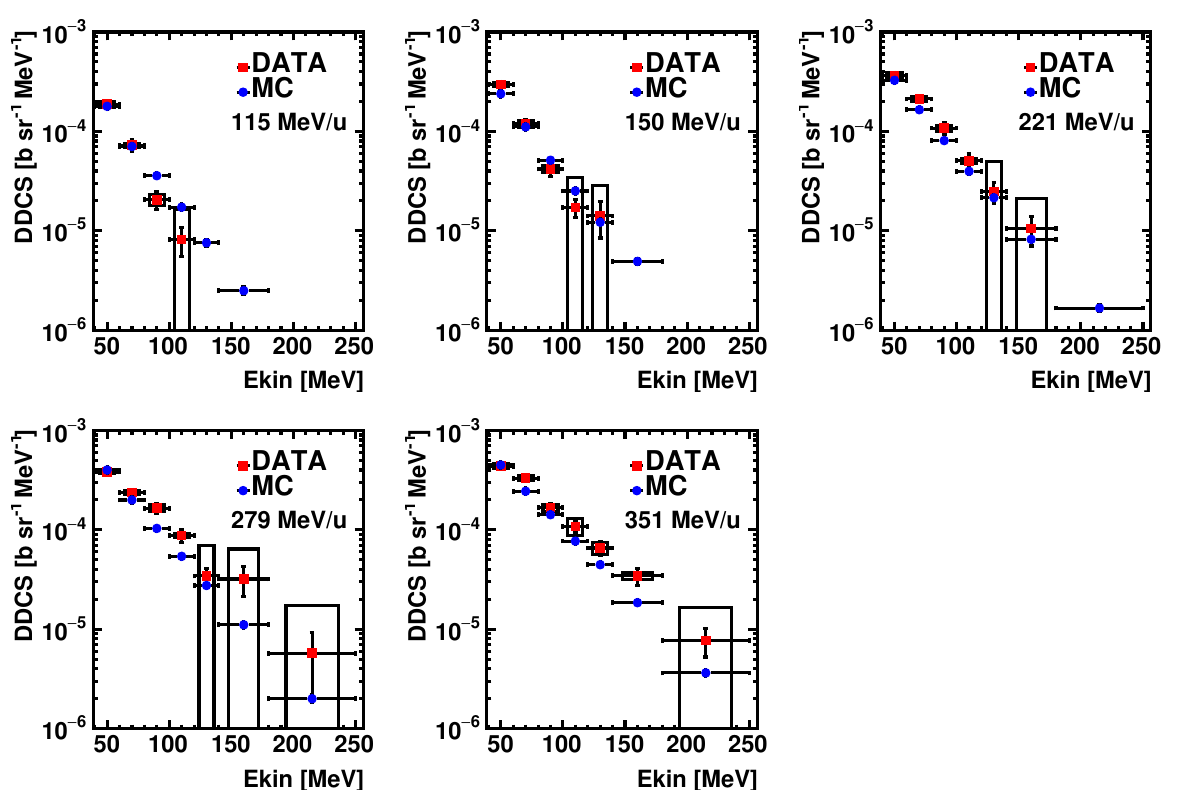}
\end{center}
\caption{Double differential cross section as a function of fragment kinetic energy for proton fragments detected at 90$\degree$, produced in the nuclear interaction of 115-351~MeV/u carbon ion beam with a graphite target. The statistical uncertainty (cross) and systematic uncertainty (empty square) on experimental data are shown as separate contributions.}
\label{fig:ddc90}
\end{figure*}

\begin{figure*}
\begin{center}
\includegraphics[width=0.8\linewidth]{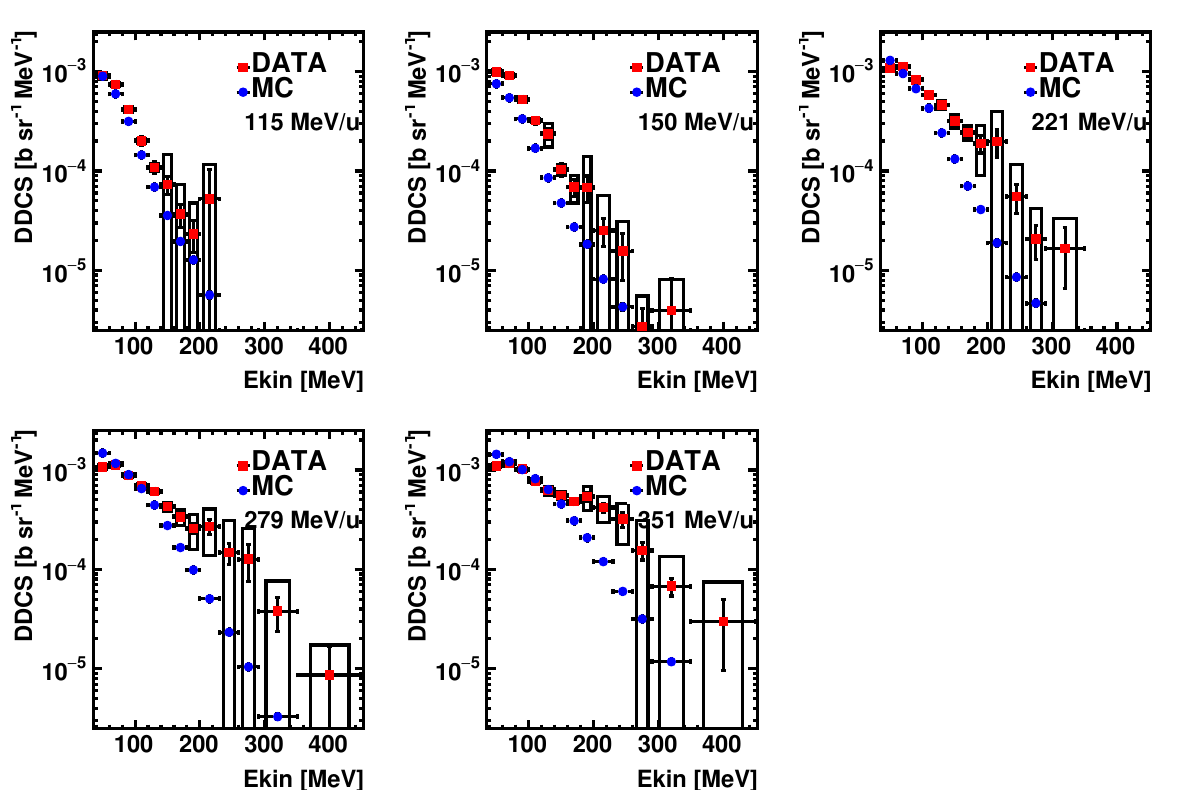}
\end{center}
\caption{Double differential cross section as a function of fragment kinetic energy for proton fragments detected at 60$\degree$, produced in the nuclear interaction of 115-351~MeV/u carbon ion beam with a graphite target. The statistical uncertainty (cross) and systematic uncertainty (empty square) on experimental data are shown as separate contributions.}
\label{fig:ddc60}
\end{figure*}

\begin{figure*}
\begin{center}
\includegraphics[width=0.8\linewidth]{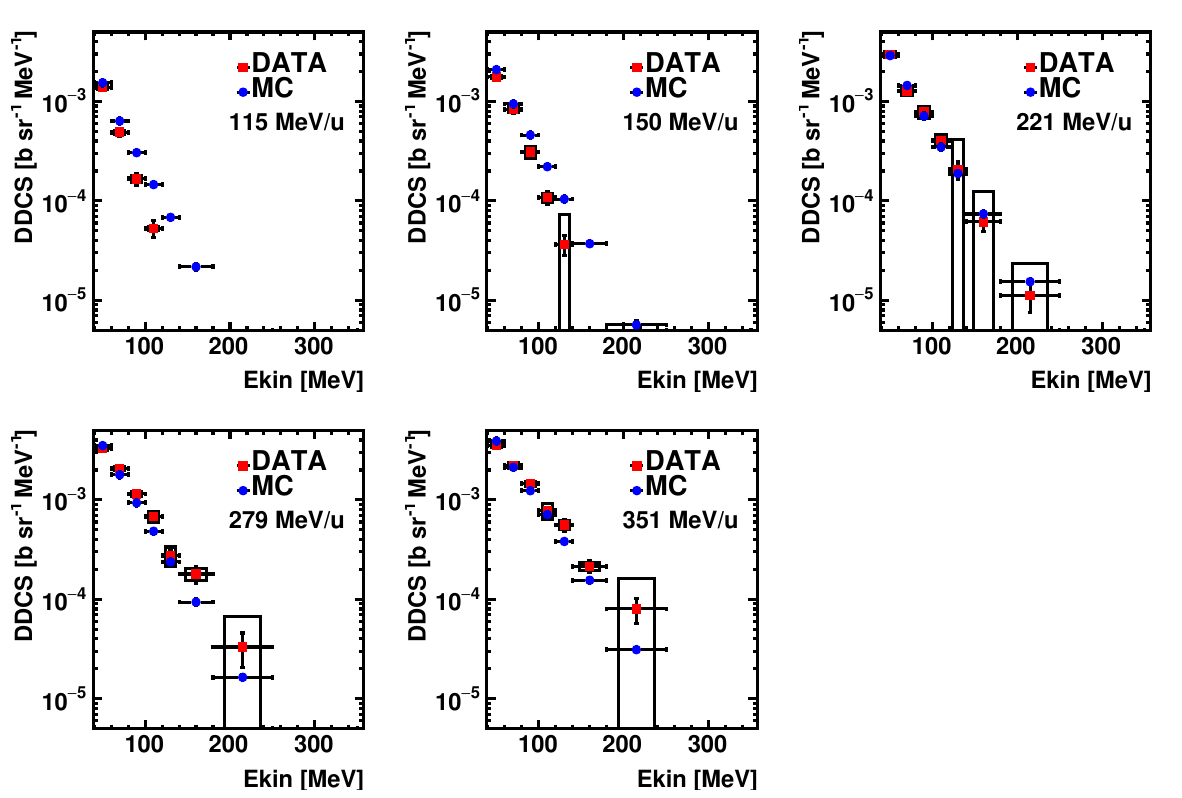}
\end{center}
\caption{Double differential cross section as a function of fragment kinetic energy for proton fragments detected at 90$\degree$, produced in the nuclear interaction of 115-351~MeV/u carbon ion beam with a polyvinyl-toluene target. The statistical uncertainty (cross) and systematic uncertainty (empty square) on experimental data are shown as separate contributions.}
\label{fig:ddej90}
\end{figure*}

\begin{figure*}
\begin{center}
\includegraphics[width=0.8\linewidth]{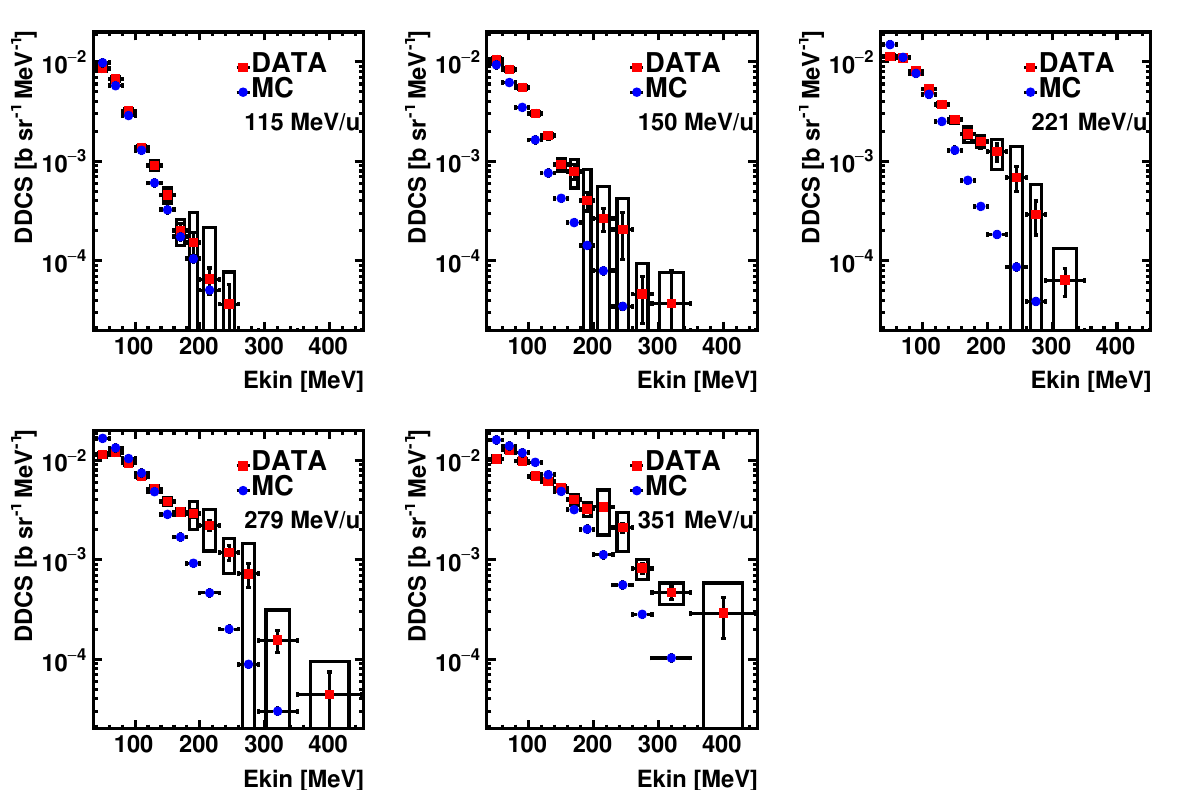}
\end{center}
\caption{Double differential cross section as a function of fragment kinetic energy for proton fragments detected at 60$\degree$, produced in the nuclear interaction of 115-351~MeV/u carbon ion beam with a polyvinyl-toluene target. The statistical uncertainty (cross) and systematic uncertainty (empty square) on experimental data are shown as separate contributions.}
\label{fig:ddej60}
\end{figure*}

\begin{figure*}
\begin{center}
\includegraphics[width=0.8\linewidth]{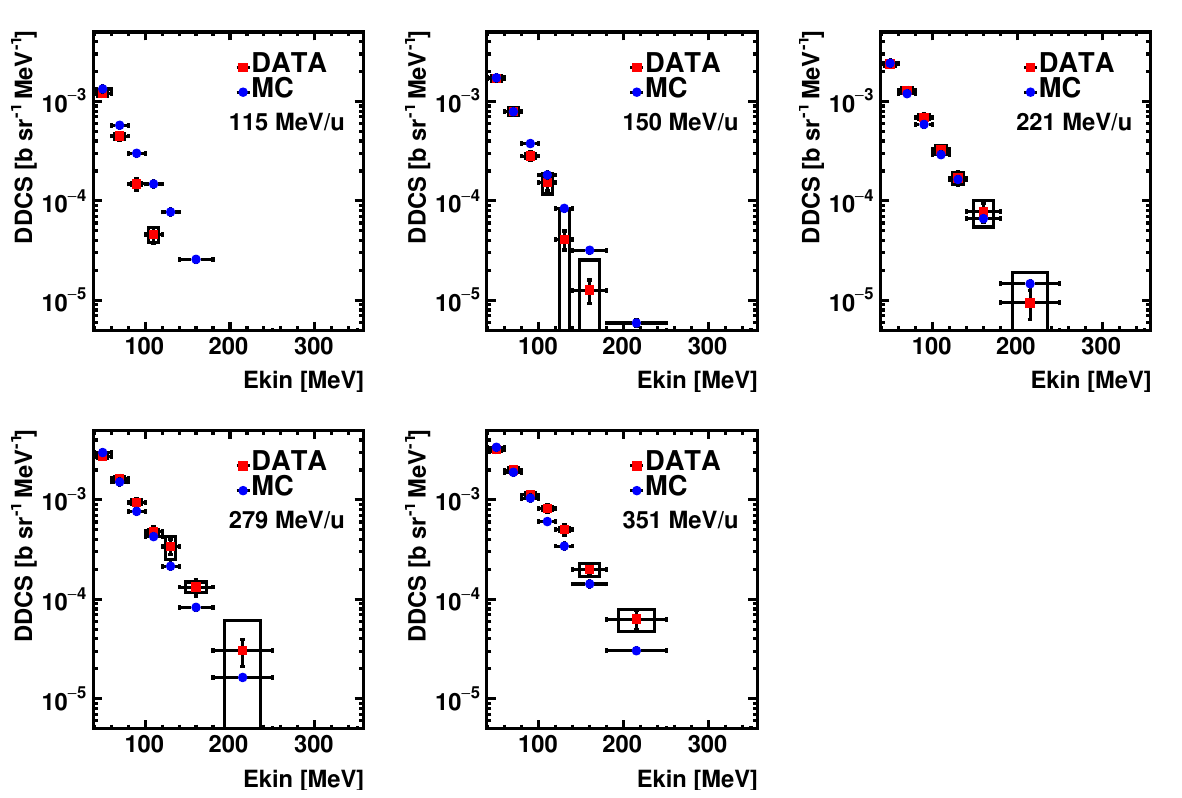}
\end{center}
\caption{Double differential cross section as a function of kinetic energy for proton fragments detected at 90$\degree$, produced in the nuclear interaction of 115-351~MeV/u carbon ion beam with a PMMA target. The statistical uncertainty (cross) and systematic uncertainty (empty square) on experimental data are shown as separate contributions.}
\label{fig:ddpmma90}
\end{figure*}

\begin{figure*}
\begin{center}
\includegraphics[width=0.8\linewidth]{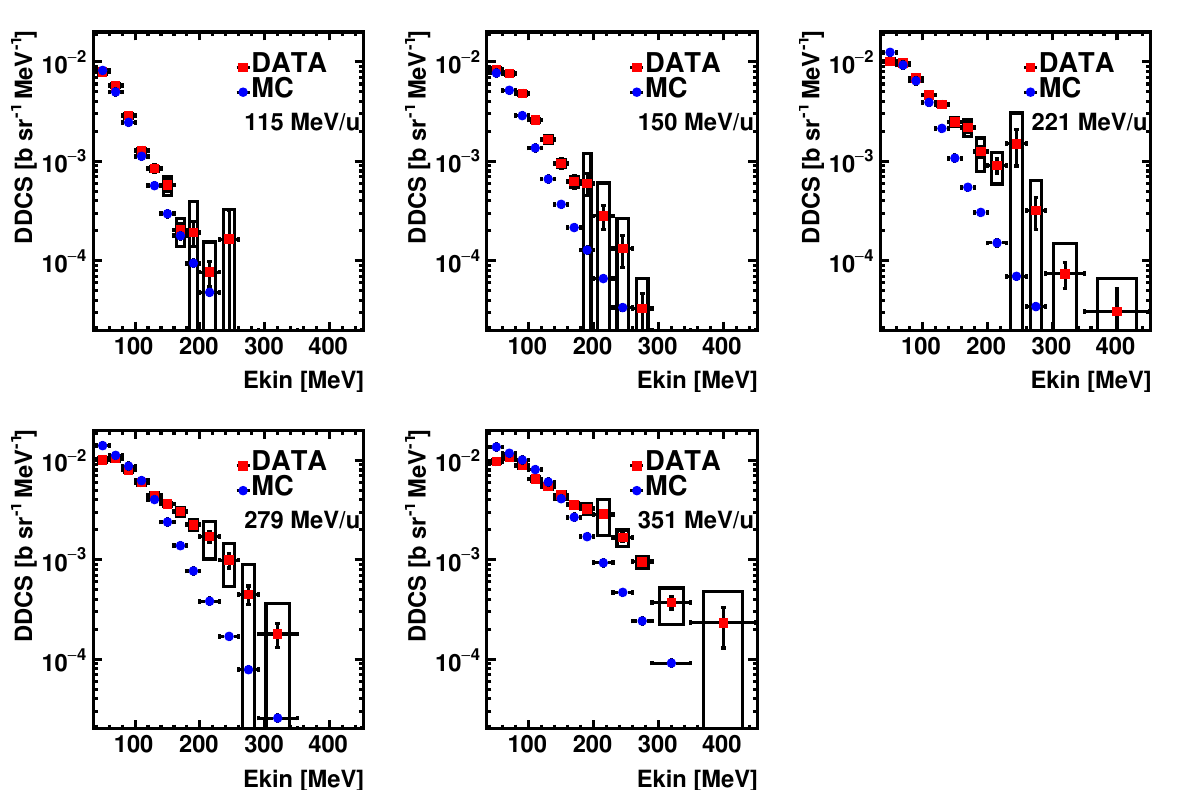}
\end{center}
\caption{Double differential cross section as a function of kinetic energy for proton fragments detected at 60$\degree$, produced in the nuclear interaction of 115-351~MeV/u carbon ion beam with a PMMA target. The statistical uncertainty (cross) and systematic uncertainty (empty square) on experimental data are shown as separate contributions.}
\label{fig:ddpmma60}
\end{figure*}

\begin{figure*}
\begin{center}
\includegraphics[width=1.\linewidth]{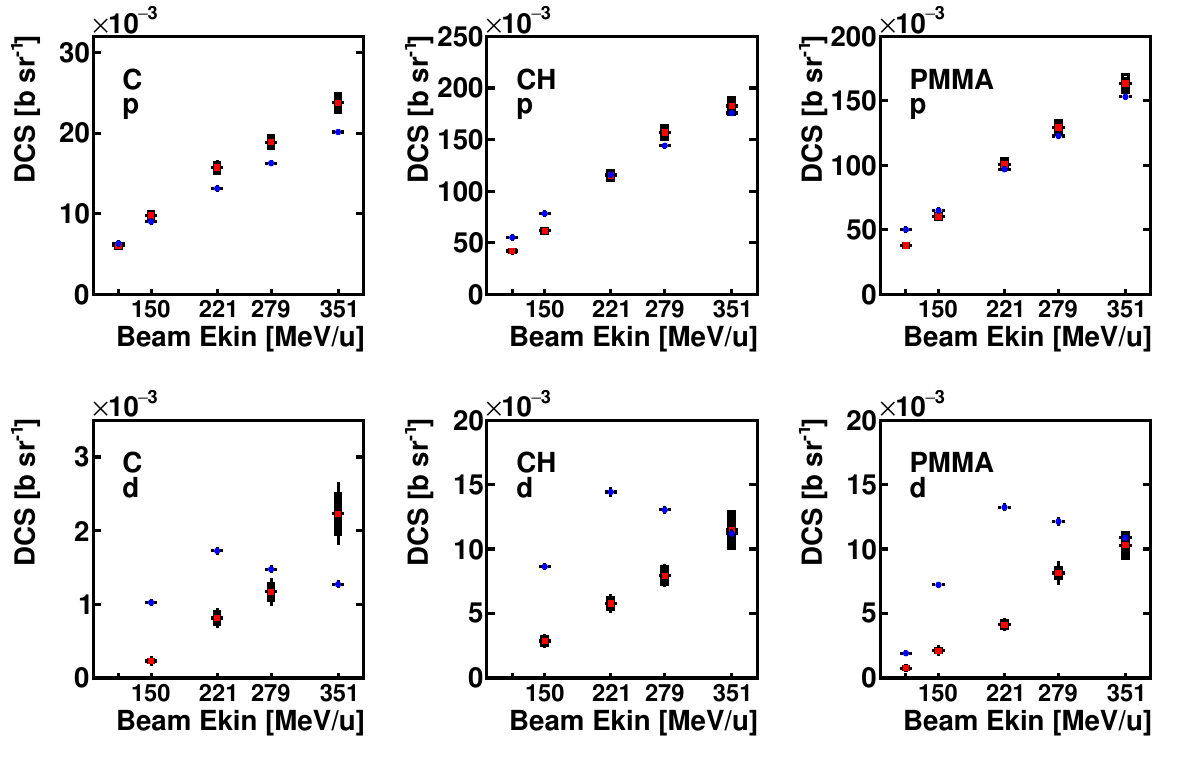}
\end{center}
\caption{Energy integrated differential cross section as a function of carbon ion beam kinetic energy for proton (top) and deuteron (bottom) fragments detected at 90$\degree$, produced in the nuclear interaction of 115-351~MeV/u carbon ion beam with graphite (C, left), polyvinyl-toluene (CH, middle) and PMMA (right) targets. Experimental data are shown as red squares, the FLUKA MC prediction is shown as blue dots. The statistical uncertainty (cross) and systematic uncertainty (empty square) on experimental data are shown as separate contributions. Cross section for tritons detected at 90$\degree$ is not reported due to insufficient statistics, as well as deuterons detected at 90$\degree$ produced by 115~MeV/u $^{12}$C ion beam impinging on C and CH targets.}
\label{fig:sdpmma90}
\end{figure*}

\begin{figure*}
\begin{center}
\includegraphics[width=1.\linewidth]{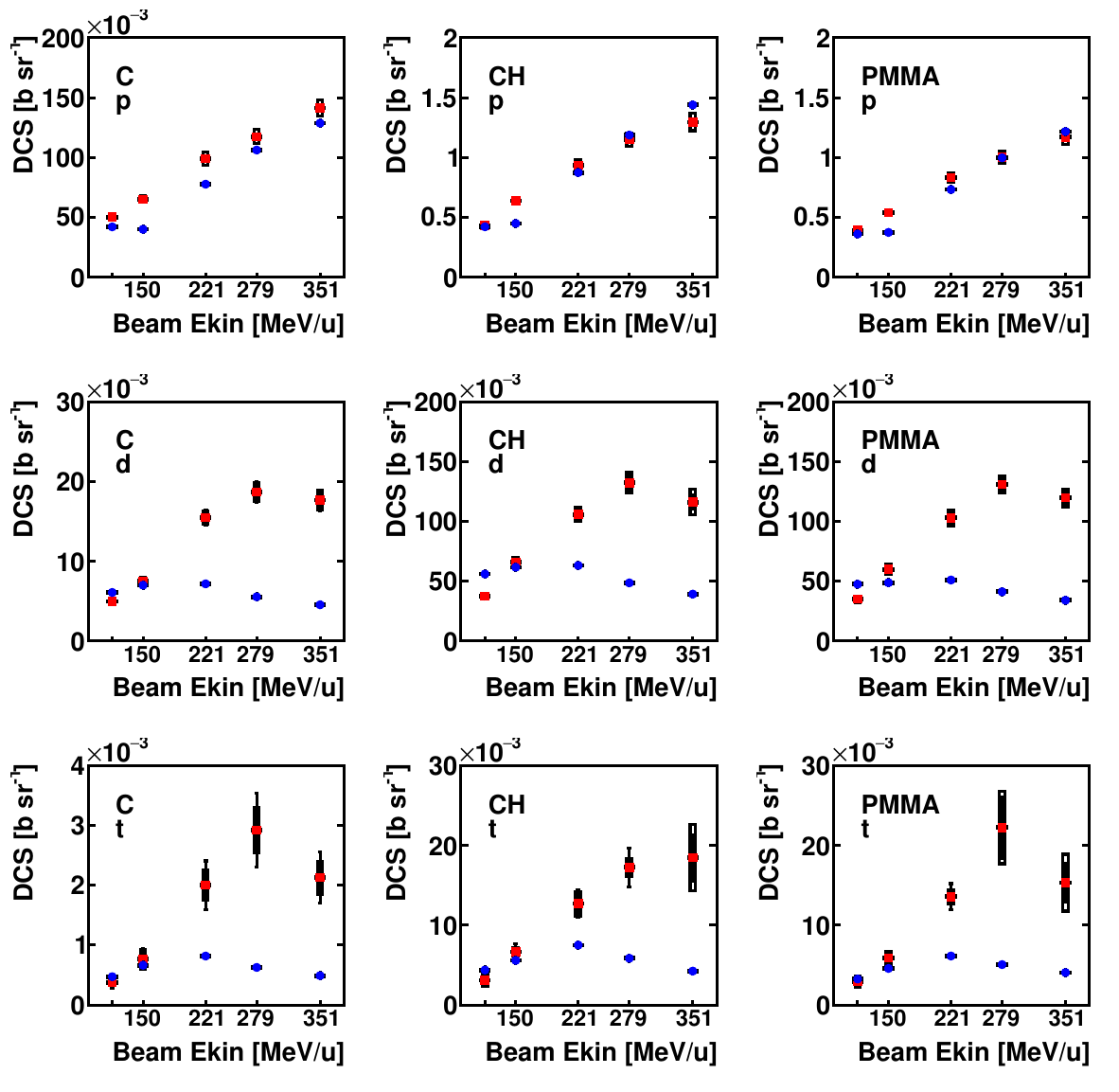}
\end{center}
\caption{Energy integrated differential cross section as a function of carbon ion beam kinetic energy for proton (top) and deuteron (bottom) fragments detected at 60$\degree$, produced in the nuclear interaction of 115-351~MeV/u carbon ion beam with graphite (C, left), polyvinyl-toluene (CH, middle) and PMMA (right) targets. Experimental data are shown as red squares, the FLUKA MC prediction is shown as blue dots. The statistical uncertainty (cross) and systematic uncertainty (empty square) on experimental data are shown as separate contributions.}
\label{fig:sdpmma60}
\end{figure*}

The results on the double differential cross section (DDCS in the following figures) of protons detected at 90$\degree$ and 60$\degree$, produced by the nuclear fragmentation of $^{12}$C ion beams of 115~-~351~MeV/u kinetic energy impinging over composite targets of C, CH and PMMA computed as described by eq.~\ref{eq:XS}, are shown as red squares, respectively, in Figs.~\ref{fig:ddc90}~-~\ref{fig:ddpmma60}. The statistical uncertainty (cross) and systematic uncertainty (empty square) on the measurements are shown as separate contributions. %Figs.~\ref{fig:sdpmma90}~-~\ref{fig:sdej60} 
The energy integrated values of the differential cross section (DCS in the following figures) are also shown as a function of the primary beam energy for the three targets (C - left panel, CH - middle panel and PMMA - right panel), for protons (p), deuterons (d) and tritons (t), detected at 90$\degree$ (Fig.~\ref{fig:sdpmma90}) and 60$\degree$ (Fig.~\ref{fig:sdpmma60}).  
The FLUKA MC prediction is superimposed to the experimental data as blue dots in all figures. Numerical values are reported in the tables shown in Appendix~\ref{sec:app}. In the case of protons, low statistics kinetic energy bins are not included in tables.

The MC prediction results on the double differential cross section are computed with the following formula:
\begin{equation}
\frac{d^2\sigma^{MC}}{d\Omega dE_{k}}(^A _ZX)= \frac{N^{trueMC}_{_Z^AX}(E_k)_{\Delta\Omega}}{N_{Y}~\Delta\Omega~\Delta E_{k}~N^{MC}_{^{12}C}}
    \label{eq:XSmc}
\end{equation}

where the numerator, shown also in eq.~\ref{eq:effidet}, is the number of generated $_Z^AX$ fragments produced in the $\Delta\Omega$ seen by the LYSO detector, $N_{Y}$, $\Delta\Omega$ and $\Delta E_{k}$ are the same quantities described in eq.~\ref{eq:XS} and $N^{MC}_{^{12}C}$ is the number of simulated carbon ions impinging the target ($\sim 10^{10}$ primary ions for each simulated target - beam configuration). The analysis strategy and in particular the MC corrections applied to the cross section formula (see eq.~\ref{eq:XS}) have been validated comparing the MC cross section, in eq.~\ref{eq:XSmc}, with the MC reconstructed one, computed following eq.~\ref{eq:XS}.

As far as deuterons and tritons are concerned, since their yield at large angle is very low, the available statistics is low and only the energy integrated differential cross sections as a function of the $^{12}$C ion beam kinetic energy in Figs.~\ref{fig:sdpmma90} and~\ref{fig:sdpmma60} can be presented. Again, statistical (cross) and systematic uncertainties (empty square) on experimental data (red squares) are shown as separate contributions, and the FLUKA MC predictions are superimposed as blue dots. %In the case of deuterons detected at 90$\degree$ produced by 115~MeV/u carbon ion primary beam impinging on C and CH targets and tritons detected at 90$\degree$, where we have the minimum yield, we could not compute such cross sections because of the insufficient statistics.

\section{\label{sec:discuss}Discussion and Conclusions}
\begin{figure*}
\begin{center}
\includegraphics[width=1.\linewidth]{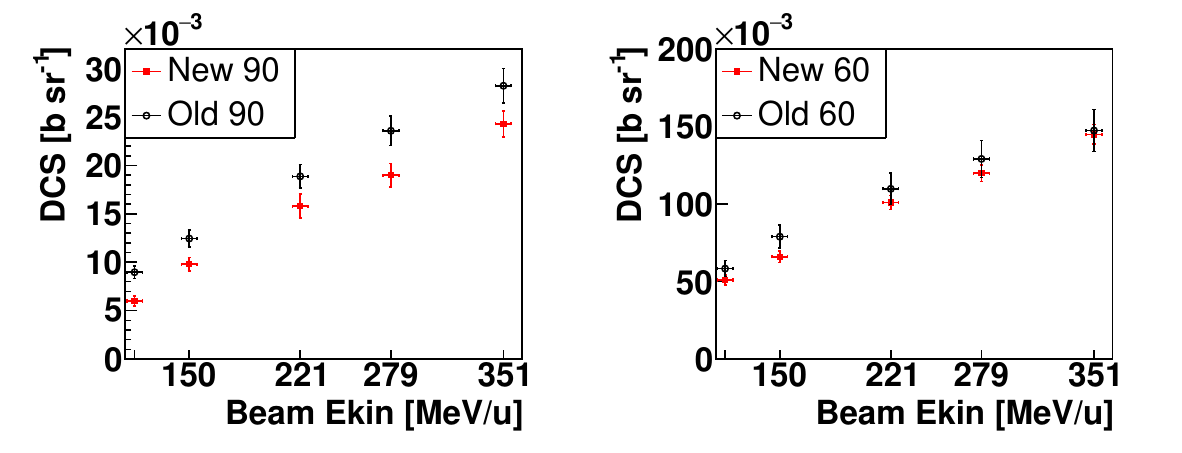}
\end{center}
\caption{Comparison of the new (red full squares) and old (black open circles) energy integrated differential cross sections obtained for secondary proton fragments (p) detected at 90$\degree$ (left) and 60$\degree$  (right) from the interaction of 115-351 MeV/u $^{12}$C ion beam with graphite (C) target.}
\label{fig:compare}
\end{figure*}

This work is devoted to the study of the emission of nucleons and light charged fragments at large polar angle in $^{12}$C collisions in the energy range used in particle therapy.
The main aim is to provide data to benchmark the models used for specific tasks, as those concerning range monitoring in ion therapy by means of the detection of light nuclear fragments emitted by the interaction of the therapeutic beam in the patient\cite{paperDP,toppi2021,fischetti2020}. 
% The main aim is to provide data to benchmark models used for specific tasks, such as range monitoring in ion therapy through the detection of light nuclear fragments emitted by the interaction of the therapeutic beam with the patient.
There have been other works in the past where data on yield at large angle from thick targets have been measured\cite{piersanti2012,marafini2017,rucinsky2018,rucinsky2019}, but no cross section measurements on thin target have been published other than ref.\cite{paperfoot}.
In the present work, the double differential cross sections of protons produced at large angles (60 and 90 degrees) from the nuclear interaction of $^{12}$C ion beams of 115-351~MeV/u impinging over graphite, polyvinyl-toluene and PMMA targets are presented, together with the energy integrated differential cross section for $^2$H and $^3$H isotopes. 
These results are obtained from a complete novel analysis of an already published dataset~\cite{paperfoot}, refining the efficiency calculation strategy, taking into account the efficiency dependency to the fragment production kinetic energy. 
Also the calculation of the fragments kinetic energy at production has been improved, applying an unfolding technique to the measured fragment kinetic energy instead of implementing an analytic correction function, as it was done in the previously published results. 
In Figure~\ref{fig:compare}, the comparison of the novel cross sections (red full squares) to the previously published ones~\cite{paperfoot} (black open circles) is shown for the case of integrated proton fragments cross sections detected at 90$\degree$ (left) and 60$\degree$ (right), produced from the interaction of carbon beams with graphite target. 
For the case of protons detected at 90$\degree$ a systematic lower shift of new results between 10-30$\%$ has been found with respect to the old data, while old and new results are in agreement at 60$\degree$ detection angle.% Similar results have been obtained for the PMMA and CH targets \textcolor{red}{da verificare!}. 

It has been also shown, for the first time, a comparison with the predictions obtained from the FLUKA MC code. 
The production of Z=1 fragments above 60 degrees represents a small fraction of their whole yield coming from the nuclear fragmentation process of $^{12}$C projectiles. 
Using FLUKA, it has been estimated that in C-C collisions at the primary energies considered in this work, the fraction of protons emitted at angles exceeding 60 degrees is about 3$\%$ of their total emission in the forward hemisphere. 
In the case of deuterons and tritons, such a percentage is even a bit lower (2.2$\%$ - 2.8$\%$).   
In the last decades, significant efforts have been made by developers to improve the reliability of phenomenological models for hadron therapy use, leading to considerable progress. 
However, it has to be remarked once again that the forward production dominates. 
The behavior of theory-driven microscopic models of high quality MC codes is generally determined by a limited number of parameters, making it challenging to achieve accurate reproduction of experimental data in the whole phase space. 
Therefore, the comparison of predictions in the large angle region can be considered a real stress test for these models, and it can be considered as a further added value of the shown results. 
Our results show that FLUKA provides reasonable predictions for protons in almost the whole explored range. 
This is instead not  for deuterons and tritons, with the exception the case of 115-150 MeV/u primary energy at 60$\degree$ and of 351 MeV/u primary energy at 90$\degree$, although in the latter case, the agreement might just be accidental, given the anomalous behaviour as a function of energy. 
For the reasons above summarized, the presence of discrepancies in this region cannot be considered surprising or alarming. 
In this respect, it is worthwhile pointing out that in the case of FLUKA, as explained in~\cite{fluka3}, there is a transition of internal models of nucleus-nucleus interactions around 125 MeV/u, from BME~\cite{bme} to rQMD~\cite{rqmd1,rqmd2}. 
The behaviour of FLUKA for deuterons and tritons for 150 MeV/u and beyond might suggest possible issues in the rQMD model near its lower energy limit. This has been recently pointed out to the FLUKA developers.
It must also be considered that, with respect to deuterons and tritons, the emission of low energy protons may be more heavily affected by later phases of the interaction, such as the pre-equilibrium process.
However, the discussion of model details it is not the aim of this work. 
Other models will be compared in the future. From the experimental point of view, it may also be important to investigate potential differences among the Z=1 isotopes measured at small polar angles. 
This task could be addressed by the FOOT experiment~\cite{footexp}.\\

\begin{acknowledgments}
%\section*{Acknowledgments}
Authors sincerely thank Marco Magi (\textit{Dipartimento di Scienze di Base e Applicate per l'Ingegneria}, Sapienza Universit\`a di Roma) for his effort in the construction of mechanical supports for the experimental setup.  We are indebted to CNAO particle therapy center for encouraging this measurement, made possible by its support and by the help of the whole CNAO staff. This work has been partially supported by the Tuscany Government, POR FSE 2014 -2020, through the PETRA - INFN-RT2 172800 Project.
\end{acknowledgments}
%
%\section*{REFERENCES}
\bibliography{CNAO.bib}% Produces the bibliography via BibTeX.
%%%%%%%%%%%%%%%%%%%%%%%%%%%%%%%%%%%%%%%%%%%%%%%%%%%%%%%%%%%%%%%%%%%%%%%%%%%%
\appendix
\section{Double Differential Cross Section Tables}
\label{sec:app}
The double differential cross sections for protons and the energy integrated differential cross sections for protons, deuterons and tritons are here reported for each carbon ion beam energy (115-351~MeV/u) and two detection arms (90$\degree$,60$\degree$).
\begin{table*}[t!]
\begin{minipage}{\columnwidth}
\begin{center}
\begin{tabular}{|c||cccc|}
\hline
\rule[-4mm]{0mm}{1cm}
 $E^p_{kin}$  & $\frac{d^2\sigma^{MC}}{d\Omega dE_k}$ & $\frac{d^2\sigma^{data}}{d\Omega dE_k}$ & $\Delta_{stat}^{data}$ & $\Delta_{sys}^{data}$\\
 $[$MeV$]$ & [b/sr/MeV] &  [b/sr/MeV] & [$\%$] & [$\%$]\\
\hline
 90$\degree$ & $\cdot 10^{-4}$  & $\cdot 10^{-4}$ & & \\
\hline
40 - 60 & 1.79 $\pm$ 0.03 & 1.9 $\pm$ 0.2 $\pm$ 0.1 & 8 & 7 \\
60 - 80 & 0.71 $\pm$ 0.02 & 0.73 $\pm$ 0.10 $\pm$ 0.04 & 13 & 5 \\
80 - 100 & 0.36 $\pm$ 0.01 & 0.21 $\pm$ 0.04 $\pm$ 0.03 & 20 & 12 \\
100 - 120 & 0.173 $\pm$ 0.008 & 0.082 $\pm$ 0.026 $\pm$ 0.009 & 32 & 11 \\
120 - 140 & 0.076 $\pm$ 0.005 & - & - & - \\
140 - 180 & 0.025 $\pm$ 0.002 & - & - & - \\
\hline
 60$\degree$ & $\cdot 10^{-4}$ & $\cdot 10^{-4}$ & & \\
\hline
40 - 60 & 8.96 $\pm$ 0.06 & 9.1 $\pm$ 0.3 $\pm$ 0.7 & 4 & 8 \\
60 - 80 & 5.95 $\pm$ 0.05 & 7.4 $\pm$ 0.4 $\pm$ 0.3 & 5 & 5 \\
80 - 100 & 3.15 $\pm$ 0.04 & 4.2 $\pm$ 0.3 $\pm$ 0.3 & 7 & 7 \\
100 - 120 & 1.45 $\pm$ 0.03 & 2.0 $\pm$ 0.2 $\pm$ 0.1 & 9 & 6 \\
120 - 140 & 0.69 $\pm$ 0.02 & 1.09 $\pm$ 0.14 $\pm$ 0.10 & 13 & 9 \\
140 - 160 & 0.36 $\pm$ 0.01 & 0.74 $\pm$ 0.15 $\pm$ 0.08 & 21 & 11 \\
160 - 180 & 0.196 $\pm$ 0.009 & 0.37 $\pm$ 0.10 $\pm$ 0.05 & 26 & 13 \\
180 - 200 & 0.128 $\pm$ 0.008 & 0.23 $\pm$ 0.08 $\pm$ 0.07 & 35 & 32 \\
200 - 230 & 0.057 $\pm$ 0.004 & 0.5 $\pm$ 0.5 $\pm$ 0.4 & 97 & 70 \\
230 - 260 & 0.021 $\pm$ 0.003 & - & - & - \\
260 - 290 & 0.008 $\pm$ 0.002 & - & - & - \\
\hline
\end{tabular}
\end{center}
\caption{Double differential cross section in kinetic energy bins at production ($E^p_{kin}$) of protons produced by 115~MeV/u $^{12}$C ion beam impinging on a C target, detected at 90$\degree$ (top panel) and 60$\degree$ (bottom panel). The production cross section from the FLUKA Monte Carlo simulation (\textit{MC}) is listed alongside the experimental cross section (\textit{data}), with the relative statistical and systematic data uncertainties reported as percentage in the last two columns. }
\end{minipage}\hfill % maximize the horizontal separation
\begin{minipage}{\columnwidth}
\begin{center}
 \begin{tabular}{|c||cccc|}
\hline
\rule[-4mm]{0mm}{1cm}
 $E^p_{kin}$  & $\frac{d^2\sigma^{MC}}{d\Omega dE_k}$ & $\frac{d^2\sigma^{data}}{d\Omega dE_k}$ & $\Delta_{stat}^{data}$ & $\Delta_{sys}^{data}$\\
 $[$MeV$]$ & [b/sr/MeV] &  [b/sr/MeV] & [$\%$] & [$\%$]\\
\hline
 90$\degree$ & $\cdot 10^{-4}$ & $\cdot 10^{-4}$ &  &  \\
\hline
40 - 60 & 2.39 $\pm$ 0.03 & 3.0 $\pm$ 0.2 $\pm$ 0.1 & 8 & 5 \\
60 - 80 & 1.11 $\pm$ 0.02 & 1.20 $\pm$ 0.13 $\pm$ 0.10 & 11 & 8 \\
80 - 100 & 0.51 $\pm$ 0.01 & 0.42 $\pm$ 0.07 $\pm$ 0.03 & 16 & 8 \\
100 - 120 & 0.252 $\pm$ 0.010 & 0.172 $\pm$ 0.036 $\pm$ 0.009 & 21 & 5 \\
120 - 140 & 0.122 $\pm$ 0.007 & 0.14 $\pm$ 0.06 $\pm$ 0.03 & 40 & 21 \\
140 - 180 & 0.049 $\pm$ 0.003 & - & - & - \\
180 - 250 & 0.0069 $\pm$ 0.0009 & - & - & - \\
\hline
 60$\degree$ & $\cdot 10^{-4}$ & $\cdot 10^{-4}$ & &  \\
\hline
40 - 60 & 7.52 $\pm$ 0.06 & 9.9 $\pm$ 0.4 $\pm$ 0.5 & 4 & 6 \\
60 - 80 & 5.44 $\pm$ 0.05 & 9.1 $\pm$ 0.4 $\pm$ 0.4 & 5 & 5 \\
80 - 100 & 3.34 $\pm$ 0.04 & 5.3 $\pm$ 0.3 $\pm$ 0.4 & 6 & 7 \\
100 - 120 & 1.69 $\pm$ 0.03 & 3.2 $\pm$ 0.3 $\pm$ 0.2 & 8 & 5 \\
120 - 140 & 0.85 $\pm$ 0.02 & 2.4 $\pm$ 0.3 $\pm$ 0.6 & 12 & 26 \\
140 - 160 & 0.48 $\pm$ 0.01 & 1.0 $\pm$ 0.1 $\pm$ 0.1 & 14 & 13 \\
160 - 180 & 0.27 $\pm$ 0.01 & 0.7 $\pm$ 0.1 $\pm$ 0.2 & 20 & 31 \\
180 - 200 & 0.184 $\pm$ 0.009 & 0.7 $\pm$ 0.2 $\pm$ 0.2 & 30 & 25 \\
200 - 230 & 0.082 $\pm$ 0.005 & 0.25 $\pm$ 0.08 $\pm$ 0.18 & 32 & 72 \\
230 - 260 & 0.043 $\pm$ 0.004 & 0.16 $\pm$ 0.08 $\pm$ 0.01 & 49 & 9 \\
260 - 290 & 0.017 $\pm$ 0.002 & 0.027 $\pm$ 0.014 $\pm$ 0.006 & 51 & 22 \\
290 - 350 & 0.0065 $\pm$ 0.0010 & 0.04 $\pm$ 0.04 $\pm$ 0.01 & 112 & 27 \\
\hline
\end{tabular}
\end{center}
\caption{Double differential cross section in kinetic energy bins at production ($E^p_{kin}$) of protons produced by 150~MeV/u $^{12}$C ion beam impinging on a C target, detected at 90$\degree$ (top panel) and 60$\degree$ (bottom panel). The production cross section from the FLUKA Monte Carlo simulation (\textit{MC}) is listed alongside the experimental cross section (\textit{data}), with the relative statistical and systematic data uncertainties reported as percentage in the last two columns. }
\end{minipage} % maximize the horizontal separation
\end{table*}

\begin{table*}[b!]
\begin{minipage}{\columnwidth}
\begin{center}
\begin{tabular}{|c||cccc|}
\hline
\rule[-4mm]{0mm}{1cm}
 $E^p_{kin}$  & $\frac{d^2\sigma^{MC}}{d\Omega dE_k}$ & $\frac{d^2\sigma^{data}}{d\Omega dE_k}$ & $\Delta_{stat}^{data}$ & $\Delta_{sys}^{data}$\\
 $[$MeV$]$ & [b/sr/MeV] &  [b/sr/MeV] & [$\%$] & [$\%$]\\
\hline
 90$\degree$ & $\cdot 10^{-4}$ & $\cdot 10^{-4}$ & &  \\
\hline
40 - 60 & 3.27 $\pm$ 0.04 & 3.6 $\pm$ 0.2 $\pm$ 0.2 & 6 & 6 \\
60 - 80 & 1.65 $\pm$ 0.03 & 2.1 $\pm$ 0.2 $\pm$ 0.1 & 9 & 7 \\
80 - 100 & 0.81 $\pm$ 0.02 & 1.08 $\pm$ 0.15 $\pm$ 0.08 & 14 & 7 \\
100 - 120 & 0.40 $\pm$ 0.01 & 0.51 $\pm$ 0.09 $\pm$ 0.03 & 18 & 6 \\
120 - 140 & 0.215 $\pm$ 0.009 & 0.25 $\pm$ 0.06 $\pm$ 0.02 & 24 & 9 \\
140 - 180 & 0.082 $\pm$ 0.004 & 0.105 $\pm$ 0.035 $\pm$ 0.007 & 33 & 7 \\
180 - 250 & 0.017 $\pm$ 0.001 & - & - & - \\
\hline
 60$\degree$ & $\cdot 10^{-4}$ & $\cdot 10^{-4}$ & &  \\
\hline
40 - 60 & 12.93 $\pm$ 0.08 & 10.9 $\pm$ 0.3 $\pm$ 0.7 & 3 & 6 \\
60 - 80 & 9.54 $\pm$ 0.07 & 11.2 $\pm$ 0.4 $\pm$ 0.7 & 4 & 7 \\
80 - 100 & 6.72 $\pm$ 0.05 & 8.2 $\pm$ 0.4 $\pm$ 0.4 & 4 & 5 \\
100 - 120 & 4.27 $\pm$ 0.04 & 5.8 $\pm$ 0.3 $\pm$ 0.4 & 6 & 6 \\
120 - 140 & 2.41 $\pm$ 0.03 & 4.6 $\pm$ 0.3 $\pm$ 0.4 & 7 & 10 \\
140 - 160 & 1.32 $\pm$ 0.02 & 3.2 $\pm$ 0.3 $\pm$ 0.5 & 10 & 16 \\
160 - 180 & 0.70 $\pm$ 0.02 & 2.5 $\pm$ 0.4 $\pm$ 0.4 & 14 & 17 \\
180 - 200 & 0.41 $\pm$ 0.01 & 1.9 $\pm$ 0.4 $\pm$ 1.0 & 20 & 52 \\
200 - 230 & 0.189 $\pm$ 0.007 & 2.0 $\pm$ 0.6 $\pm$ 0.4 & 31 & 19 \\
230 - 260 & 0.086 $\pm$ 0.005 & 0.6 $\pm$ 0.2 $\pm$ 0.3 & 33 & 47 \\
260 - 290 & 0.047 $\pm$ 0.004 & 0.21 $\pm$ 0.08 $\pm$ 0.05 & 38 & 23 \\
290 - 350 & 0.017 $\pm$ 0.002 & 0.17 $\pm$ 0.10 $\pm$ 0.03 & 61 & 17 \\
350 - 450 & 0.0028 $\pm$ 0.0005 & - & - & - \\
\hline
\end{tabular}
\end{center}
\caption{Double differential cross section in kinetic energy bins at production ($E^p_{kin}$) of protons produced by 221~MeV/u $^{12}$C ion beam impinging on a C target, detected at 90$\degree$ (top panel) and 60$\degree$ (bottom panel). The production cross section from the FLUKA Monte Carlo simulation (\textit{MC}) is listed alongside the experimental cross section (\textit{data}), with the relative statistical and systematic data uncertainties reported as percentage in the last two columns. }
\end{minipage}\hfill % maximize the horizontal separation
\begin{minipage}{\columnwidth}
\begin{center}
\begin{tabular}{|c||cccc|}
\hline
\rule[-4mm]{0mm}{1cm}
 $E^p_{kin}$  & $\frac{d^2\sigma^{MC}}{d\Omega dE_k}$ & $\frac{d^2\sigma^{data}}{d\Omega dE_k}$ & $\Delta_{stat}^{data}$ & $\Delta_{sys}^{data}$\\
 $[$MeV$]$ & [b/sr/MeV] &  [b/sr/MeV] & [$\%$] & [$\%$]\\
\hline
 90$\degree$ & $\cdot 10^{-4}$ & $\cdot 10^{-4}$ & & \\
\hline
40 - 60 & 3.99 $\pm$ 0.04 & 3.8 $\pm$ 0.2 $\pm$ 0.2 & 5 & 4 \\
60 - 80 & 2.00 $\pm$ 0.03 & 2.4 $\pm$ 0.2 $\pm$ 0.1 & 8 & 5 \\
80 - 100 & 1.03 $\pm$ 0.02 & 1.6 $\pm$ 0.2 $\pm$ 0.1 & 12 & 8 \\
100 - 120 & 0.54 $\pm$ 0.01 & 0.88 $\pm$ 0.14 $\pm$ 0.04 & 15 & 5 \\
120 - 140 & 0.28 $\pm$ 0.01 & 0.34 $\pm$ 0.06 $\pm$ 0.03 & 18 & 9 \\
140 - 180 & 0.111 $\pm$ 0.005 & 0.32 $\pm$ 0.11 $\pm$ 0.02 & 34 & 7 \\
180 - 250 & 0.020 $\pm$ 0.001 & 0.06 $\pm$ 0.04 $\pm$ 0.10 & 62 & 177 \\
\hline
 60$\degree$ & $\cdot 10^{-4}$  & $\cdot 10^{-4}$ & & \\
\hline
40 - 60 & 14.77 $\pm$ 0.08 & 10.7 $\pm$ 0.3 $\pm$ 0.7 & 3 & 6 \\
60 - 80 & 11.65 $\pm$ 0.07 & 11.3 $\pm$ 0.4 $\pm$ 0.5 & 3 & 5 \\
80 - 100 & 8.98 $\pm$ 0.06 & 8.8 $\pm$ 0.3 $\pm$ 0.5 & 4 & 5 \\
100 - 120 & 6.50 $\pm$ 0.05 & 6.9 $\pm$ 0.3 $\pm$ 0.4 & 4 & 5 \\
120 - 140 & 4.44 $\pm$ 0.04 & 6.1 $\pm$ 0.3 $\pm$ 0.3 & 6 & 5 \\
140 - 160 & 2.76 $\pm$ 0.04 & 4.3 $\pm$ 0.3 $\pm$ 0.4 & 7 & 10 \\
160 - 180 & 1.66 $\pm$ 0.03 & 3.4 $\pm$ 0.3 $\pm$ 0.6 & 9 & 17 \\
180 - 200 & 0.99 $\pm$ 0.02 & 2.6 $\pm$ 0.3 $\pm$ 1.0 & 12 & 39 \\
200 - 230 & 0.51 $\pm$ 0.01 & 2.7 $\pm$ 0.5 $\pm$ 1.3 & 17 & 50 \\
230 - 260 & 0.234 $\pm$ 0.008 & 1.5 $\pm$ 0.3 $\pm$ 1.0 & 23 & 70 \\
260 - 290 & 0.105 $\pm$ 0.006 & 1.3 $\pm$ 0.5 $\pm$ 0.7 & 41 & 55 \\
290 - 350 & 0.033 $\pm$ 0.002 & 0.4 $\pm$ 0.1 $\pm$ 0.2 & 37 & 51 \\
350 - 450 & 0.0053 $\pm$ 0.0007 & 0.09 $\pm$ 0.08 $\pm$ 0.04 & 93 & 48 \\
\hline
\end{tabular}
\end{center}
\caption{Double differential cross section in kinetic energy bins at production ($E^p_{kin}$) of protons produced by 279~MeV/u $^{12}$C ion beam impinging on a C target, detected at 90$\degree$ (top panel) and 60$\degree$ (bottom panel). The production cross section from the FLUKA Monte Carlo simulation (\textit{MC}) is listed alongside the experimental cross section (\textit{data}), with the relative statistical and systematic data uncertainties reported as percentage in the last two columns. }
\end{minipage} % maximize the horizontal separation
\end{table*}

\begin{table*}[h]
\begin{minipage}{\columnwidth}
\begin{center}
\begin{tabular}{|c||cccc|}
\hline
\rule[-4mm]{0mm}{1cm}
$E^p_{kin}$  & $\frac{d^2\sigma^{MC}}{d\Omega dE_k}$ & $\frac{d^2\sigma^{data}}{d\Omega dE_k}$ & $\Delta_{stat}^{data}$ & $\Delta_{sys}^{data}$\\
$[$MeV$]$ & [b/sr/MeV] &  [b/sr/MeV] & [$\%$] & [$\%$]\\
\hline
 90$\degree$ & $\cdot 10^{-4}$ & $\cdot 10^{-4}$ &  & \\
\hline
40 - 60 & 4.49 $\pm$ 0.04 & 4.4 $\pm$ 0.2 $\pm$ 0.3 & 5 & 6 \\
60 - 80 & 2.44 $\pm$ 0.03 & 3.3 $\pm$ 0.2 $\pm$ 0.2 & 7 & 5 \\
80 - 100 & 1.42 $\pm$ 0.02 & 1.7 $\pm$ 0.2 $\pm$ 0.1 & 9 & 8 \\
100 - 120 & 0.77 $\pm$ 0.02 & 1.1 $\pm$ 0.1 $\pm$ 0.2 & 13 & 19 \\
120 - 140 & 0.45 $\pm$ 0.01 & 0.66 $\pm$ 0.11 $\pm$ 0.09 & 16 & 14 \\
140 - 180 & 0.185 $\pm$ 0.006 & 0.34 $\pm$ 0.07 $\pm$ 0.03 & 19 & 7 \\
180 - 250 & 0.036 $\pm$ 0.002 & 0.08 $\pm$ 0.02 $\pm$ 0.04 & 32 & 57 \\
250 - 350 & 0.0027 $\pm$ 0.0005 & - & - & - \\
\hline
 60$\degree$ & $\cdot 10^{-4}$  & $\cdot 10^{-4}$ &  &  \\
\hline
40 - 60 & 14.38 $\pm$ 0.08 & 11.0 $\pm$ 0.3 $\pm$ 0.5 & 3 & 5 \\
60 - 80 & 12.11 $\pm$ 0.07 & 11.6 $\pm$ 0.4 $\pm$ 0.6 & 3 & 5 \\
80 - 100 & 10.08 $\pm$ 0.07 & 10.2 $\pm$ 0.4 $\pm$ 0.5 & 4 & 5 \\
100 - 120 & 8.17 $\pm$ 0.06 & 7.7 $\pm$ 0.3 $\pm$ 0.4 & 4 & 5 \\
120 - 140 & 6.35 $\pm$ 0.05 & 6.2 $\pm$ 0.3 $\pm$ 0.6 & 4 & 10 \\
140 - 160 & 4.54 $\pm$ 0.04 & 5.5 $\pm$ 0.3 $\pm$ 0.5 & 6 & 8 \\
160 - 180 & 3.09 $\pm$ 0.04 & 4.8 $\pm$ 0.3 $\pm$ 0.2 & 7 & 5 \\
180 - 200 & 2.08 $\pm$ 0.03 & 5.4 $\pm$ 0.5 $\pm$ 1.5 & 10 & 27 \\
200 - 230 & 1.20 $\pm$ 0.02 & 4.2 $\pm$ 0.5 $\pm$ 1.2 & 11 & 29 \\
230 - 260 & 0.60 $\pm$ 0.01 & 3.2 $\pm$ 0.6 $\pm$ 1.4 & 17 & 44 \\
260 - 290 & 0.317 $\pm$ 0.010 & 1.6 $\pm$ 0.3 $\pm$ 0.6 & 20 & 38 \\
290 - 350 & 0.118 $\pm$ 0.004 & 0.7 $\pm$ 0.1 $\pm$ 0.3 & 20 & 38 \\
350 - 450 & 0.021 $\pm$ 0.001 & 0.3 $\pm$ 0.2 $\pm$ 0.4 & 68 & 118 \\
450 - 650 & 0.0012 $\pm$ 0.0002 & - & - & - \\
\hline
\end{tabular}
\end{center}
\caption{Double differential cross section in kinetic energy bins at production ($E^p_{kin}$) of protons produced by 351~MeV/u $^{12}$C ion beam impinging on a C target, detected at 90$\degree$ (top panel) and 60$\degree$ (bottom panel). The production cross section from the FLUKA Monte Carlo simulation (\textit{MC}) is listed alongside the experimental cross section (\textit{data}), with the relative statistical and systematic data uncertainties reported as percentage in the last two columns. }
\end{minipage}
\begin{minipage}{\columnwidth}
\begin{center}
\begin{tabular}{|c||cccc|}
\hline
\rule[-4mm]{0mm}{1cm}
 $E^C_{kin}$ & $\frac{d\sigma^{MC}}{d\Omega}$ &  $\frac{d\sigma^{data}}{d\Omega}$ & $\Delta_{stat}^{data}$ & $\Delta_{sys}^{data}$ \\
 $[$MeV/u$]$  & [b/sr] & [b/sr] & [$\%$] &  [$\%$]\\
\hline
 90$\degree$ & $\cdot 10^{-3}$ & $\cdot 10^{-3}$  &  &  \\
\hline
115 & 6.31 $\pm$ 0.07 & 6.0 $\pm$ 0.4 $\pm$ 0.3 & 6 & 5 \\
150 & 9.01 $\pm$ 0.08 & 9.8 $\pm$ 0.5 $\pm$ 0.5 & 5 & 5 \\
221 & 13.1 $\pm$ 0.1 & 15.7 $\pm$ 0.7 $\pm$ 0.7 & 4 & 4 \\
279 & 16.3 $\pm$ 0.1 & 18.9 $\pm$ 0.7 $\pm$ 0.8 & 4 & 4 \\
351 & 20.1 $\pm$ 0.1 & 23.8 $\pm$ 0.8 $\pm$ 1.2 & 4 & 5 \\
\hline
 60$\degree$ & $\cdot 10^{-2}$ & $\cdot 10^{-2}$ &  &  \\
\hline
115 & 4.20 $\pm$ 0.02 & 5.0 $\pm$ 0.1 $\pm$ 0.2 & 3 & 5 \\
150 & 4.00 $\pm$ 0.02 & 6.5 $\pm$ 0.2 $\pm$ 0.3 & 2 & 4 \\
221 & 7.77 $\pm$ 0.03 & 9.9 $\pm$ 0.2 $\pm$ 0.5 & 2 & 5 \\
279 & 10.62 $\pm$ 0.03 & 11.7 $\pm$ 0.2 $\pm$ 0.6 & 2 & 5 \\
351 & 12.87 $\pm$ 0.03 & 14.1 $\pm$ 0.2 $\pm$ 0.7 & 1 & 5 \\
\hline
\end{tabular}
\end{center}
\caption{Energy integrated differential cross section of protons produced by $^{12}$C ion beam of kinetic energy ($E^C_{kin}$) ranging from 115~MeV/u to 351~MeV/u impinging on a C target. Protons are detected at 90$\degree$ (top panel) and 60$\degree$ (bottom panel). The production cross section from the FLUKA Monte Carlo simulation (\textit{MC}) is listed alongside the experimental cross section (\textit{data}), with the relative statistical and systematic data uncertainties reported as percentage in the last two columns.}
\end{minipage}
\end{table*}

\begin{table*}[h]
\begin{minipage}{\columnwidth}
\begin{center}
\begin{tabular}{|c || c  c  c  c |}
\hline
\rule[-4mm]{0mm}{1cm}
 $E^C_{kin}$ & $\frac{d\sigma^{MC}}{d\Omega}$ &  $\frac{d\sigma^{data}}{d\Omega}$ & $\Delta_{stat}^{data}$ & $\Delta_{sys}^{data}$ \\
 $[$MeV/u$]$  & [b/sr] & [b/sr] & [$\%$] &  [$\%$]\\
\hline
 90$\degree$ & $\cdot 10^{-3}$& $\cdot 10^{-3}$  & & \\
\hline
150 & 1.02 $\pm$ 0.03 & 0.23 $\pm$ 0.05 $\pm$ 0.02 & 20 & 10 \\
221 & 1.72 $\pm$ 0.04 & 0.81 $\pm$ 0.12 $\pm$ 0.09 & 15 & 12 \\
279 & 1.47 $\pm$ 0.03 & 1.2 $\pm$ 0.2 $\pm$ 0.1 & 15 & 10 \\
351 & 1.27 $\pm$ 0.03 & 2.2 $\pm$ 0.4 $\pm$ 0.3 & 18 & 12 \\
\hline
 60$\degree$ & $\cdot 10^{-2}$ & $\cdot 10^{-2}$ & & \\
\hline
115 & 0.609 $\pm$ 0.007 & 0.50 $\pm$ 0.03 $\pm$ 0.03 & 7 & 7 \\
150 & 0.700 $\pm$ 0.008 & 0.75 $\pm$ 0.04 $\pm$ 0.05 & 6 & 6 \\
221 & 0.717 $\pm$ 0.008 & 1.55 $\pm$ 0.09 $\pm$ 0.08 & 6 & 5 \\
279 & 0.555 $\pm$ 0.007 & 1.9 $\pm$ 0.1 $\pm$ 0.1 & 7 & 6 \\
351 & 0.455 $\pm$ 0.006 & 1.8 $\pm$ 0.1 $\pm$ 0.1 & 7 & 7 \\
\hline
\end{tabular}
\end{center}
\caption{Energy integrated differential cross section of deuterons produced by $^{12}$C ion beam of kinetic energy ($E^C_{kin}$) ranging from 115~MeV/u to 351~MeV/u impinging on a C target. Deuterons are detected at 90$\degree$ (top panel) and 60$\degree$ (bottom panel). The production cross section from the FLUKA Monte Carlo simulation (\textit{MC}) is listed alongside the experimental cross section (\textit{data}), with the relative statistical and systematic data uncertainties reported as percentage in the last two columns. Cross section of 115~MeV/u $^{12}$C ion beam impinging on a C target for deuterons detected at 90$\degree$ is not reported due to insufficient statistics.}
\end{minipage}
\begin{minipage}{\columnwidth}
\begin{center}
\begin{tabular}{ |c || c  c  c  c| }
\hline
\rule[-4mm]{0mm}{1cm}
 $E^C_{kin}$ & $\frac{d\sigma^{MC}}{d\Omega}$ &  $\frac{d\sigma^{data}}{d\Omega}$ & $\Delta_{stat}^{data}$ & $\Delta_{sys}^{data}$ \\
 $[$MeV/u$]$  & [b/sr] & [b/sr] & [$\%$] &  [$\%$]\\
\hline
 60$\degree$ & $\cdot 10^{-3}$ & $\cdot 10^{-3}$ & & \\
\hline
115 & 0.47 $\pm$ 0.02 & 0.38 $\pm$ 0.09 $\pm$ 0.05 & 25 & 14 \\
150 & 0.66 $\pm$ 0.02 & 0.8 $\pm$ 0.2 $\pm$ 0.2 & 23 & 21 \\
221 & 0.82 $\pm$ 0.03 & 2.0 $\pm$ 0.4 $\pm$ 0.2 & 20 & 12 \\
279 & 0.63 $\pm$ 0.02 & 2.9 $\pm$ 0.6 $\pm$ 0.4 & 21 & 13 \\
351 & 0.49 $\pm$ 0.02 & 2.1 $\pm$ 0.4 $\pm$ 0.3 & 20 & 13 \\
\hline
\end{tabular}
\end{center}
\caption{Energy integrated differential cross section of tritons produced by $^{12}$C ion beam of kinetic energy ($E^C_{kin}$) ranging from 115~MeV/u to 351~MeV/u impinging on a C target. Tritons are detected at 60$\degree$. The production cross section from the FLUKA Monte Carlo simulation (\textit{MC}) is listed alongside the experimental cross section (\textit{data}), with the relative statistical and systematic data uncertainties reported as percentage in the last two columns. Cross section for tritons detected at 90$\degree$ is not reported due to insufficient statistics.}
\end{minipage}
\end{table*}

\begin{table*}
\begin{minipage}{\columnwidth}
\begin{center}
\begin{tabular}{|c || c  c  c  c |}
\hline
\rule[-4mm]{0mm}{1cm}
 $E^p_{kin}$  & $\frac{d^2\sigma^{MC}}{d\Omega dE_k}$ & $\frac{d^2\sigma^{data}}{d\Omega dE_k}$ & $\Delta_{stat}^{data}$ & $\Delta_{sys}^{data}$\\
 $[$MeV$]$ & [b/sr/MeV] &  [b/sr/MeV] & [$\%$] & [$\%$]\\
\hline
 90$\degree$ & $\cdot 10^{-5}$  & $\cdot 10^{-5}$ & & \\
\hline
40 - 60 & 154 $\pm$ 2 & 138 $\pm$ 8 $\pm$ 6 & 6 & 4 \\
60 - 80 & 64 $\pm$ 1 & 49 $\pm$ 5 $\pm$ 2 & 10 & 4 \\
80 - 100 & 30.6 $\pm$ 0.7 & 17 $\pm$ 2 $\pm$ 1 & 14 & 7 \\
100 - 120 & 14.6 $\pm$ 0.5 & 5.3 $\pm$ 1.0 $\pm$ 0.2 & 19 & 4 \\
120 - 140 & 6.8 $\pm$ 0.3 & - & - & - \\
140 - 180 & 2.2 $\pm$ 0.1 & - & - & - \\
180 - 250 & 0.23 $\pm$ 0.03 & - & - & - \\
\hline
 60$\degree$ & $\cdot 10^{-5}$ & $\cdot 10^{-5}$ & & \\
\hline
40 - 60 & 979 $\pm$ 4 & 857 $\pm$ 21 $\pm$ 54 & 2 & 6 \\
60 - 80 & 576 $\pm$ 3 & 671 $\pm$ 21 $\pm$ 33 & 3 & 5 \\
80 - 100 & 289 $\pm$ 2 & 320 $\pm$ 14 $\pm$ 18 & 4 & 5 \\
100 - 120 & 129 $\pm$ 2 & 137 $\pm$ 9 $\pm$ 6 & 6 & 4 \\
120 - 140 & 61 $\pm$ 1 & 91 $\pm$ 9 $\pm$ 10 & 10 & 11 \\
140 - 160 & 32.6 $\pm$ 0.8 & 46 $\pm$ 6 $\pm$ 8 & 13 & 18 \\
160 - 180 & 17.4 $\pm$ 0.6 & 20 $\pm$ 3 $\pm$ 6 & 17 & 29 \\
180 - 200 & 10.5 $\pm$ 0.5 & 15 $\pm$ 4 $\pm$ 4 & 27 & 26 \\
200 - 230 & 5.1 $\pm$ 0.3 & 7 $\pm$ 2 $\pm$ 14 & 29 & 212 \\
230 - 260 & 1.6 $\pm$ 0.1 & 4 $\pm$ 2 $\pm$ 2 & 57 & 53 \\
260 - 290 & 0.71 $\pm$ 0.10 & - & - & - \\
\hline
\end{tabular}
\end{center}
\caption{Double differential cross section in kinetic energy bins at production ($E^p_{kin}$) of protons produced by 115~MeV/u $^{12}$C ion beam impinging on a CH target, detected at 90$\degree$ (top panel) and 60$\degree$ (bottom panel). The production cross section from the FLUKA Monte Carlo simulation (\textit{MC}) is listed alongside the experimental cross section (\textit{data}), with the relative statistical and systematic data uncertainties reported as percentage in the last two columns. }
\end{minipage}
\begin{minipage}{\columnwidth}
\begin{center}
\begin{tabular}{|c||cccc|}
\hline
\rule[-4mm]{0mm}{1cm}
 $E^p_{kin}$  & $\frac{d^2\sigma^{MC}}{d\Omega dE_k}$ & $\frac{d^2\sigma^{data}}{d\Omega dE_k}$ & $\Delta_{stat}^{data}$ & $\Delta_{sys}^{data}$\\
 $[$MeV$]$ & [b/sr/MeV] &  [b/sr/MeV] & [$\%$] & [$\%$]\\
\hline
 90$\degree$ & $\cdot 10^{-5}$  & $\cdot 10^{-5}$ & &  \\
\hline
40 - 60 & 209 $\pm$ 2 & 176 $\pm$ 9 $\pm$ 7 & 5 & 4 \\
60 - 80 & 95 $\pm$ 1 & 83 $\pm$ 7 $\pm$ 6 & 8 & 7 \\
80 - 100 & 45.9 $\pm$ 0.9 & 31 $\pm$ 3 $\pm$ 4 & 11 & 14 \\
100 - 120 & 22.1 $\pm$ 0.6 & 11 $\pm$ 2 $\pm$ 1 & 15 & 10 \\
120 - 140 & 10.4 $\pm$ 0.4 & 3.7 $\pm$ 0.8 $\pm$ 0.4 & 23 & 12 \\
140 - 180 & 3.7 $\pm$ 0.2 & - & - & - \\
180 - 250 & 0.57 $\pm$ 0.05 & - & - & - \\
\hline
 60$\degree$ & $\cdot 10^{-5}$  & $\cdot 10^{-5}$ & &  \\
\hline
40 - 60 & 934 $\pm$ 4 & 1059 $\pm$ 27 $\pm$ 53 & 3 & 5 \\
60 - 80 & 620 $\pm$ 4 & 843 $\pm$ 25 $\pm$ 39 & 3 & 5 \\
80 - 100 & 349 $\pm$ 3 & 556 $\pm$ 23 $\pm$ 28 & 4 & 5 \\
100 - 120 & 164 $\pm$ 2 & 304 $\pm$ 18 $\pm$ 15 & 6 & 5 \\
120 - 140 & 76 $\pm$ 1 & 181 $\pm$ 15 $\pm$ 8 & 8 & 5 \\
140 - 160 & 42.5 $\pm$ 0.9 & 93 $\pm$ 10 $\pm$ 14 & 11 & 15 \\
160 - 180 & 24.2 $\pm$ 0.7 & 79 $\pm$ 14 $\pm$ 25 & 17 & 32 \\
180 - 200 & 14.3 $\pm$ 0.5 & 40 $\pm$ 9 $\pm$ 18 & 21 & 44 \\
200 - 230 & 7.9 $\pm$ 0.3 & 27 $\pm$ 7 $\pm$ 15 & 26 & 55 \\
230 - 260 & 3.5 $\pm$ 0.2 & 21 $\pm$ 10 $\pm$ 9 & 50 & 45 \\
260 - 290 & 1.7 $\pm$ 0.2 & 5 $\pm$ 2 $\pm$ 2 & 49 & 41 \\
290 - 350 & 0.45 $\pm$ 0.05 & 4 $\pm$ 4 $\pm$ 2 & 113 & 42 \\
\hline
\end{tabular}
\end{center}
\caption{Double differential cross section in kinetic energy bins at production ($E^p_{kin}$) of protons produced by 150~MeV/u $^{12}$C ion beam impinging on a CH target, detected at 90$\degree$ (top panel) and 60$\degree$ (bottom panel). The production cross section from the FLUKA Monte Carlo simulation (\textit{MC}) is listed alongside the experimental cross section (\textit{data}), with the relative statistical and systematic data uncertainties reported as percentage in the last two columns. }
\end{minipage}
\end{table*}

\begin{table*}
\begin{minipage}{\columnwidth}
\begin{center}
\begin{tabular}{|c||cccc|}
\hline
\rule[-4mm]{0mm}{1cm}
 $E^p_{kin}$  & $\frac{d^2\sigma^{MC}}{d\Omega dE_k}$ & $\frac{d^2\sigma^{data}}{d\Omega dE_k}$ & $\Delta_{stat}^{data}$ & $\Delta_{sys}^{data}$\\
 $[$MeV$]$ & [b/sr/MeV] &  [b/sr/MeV] & [$\%$] & [$\%$]\\
\hline
 90$\degree$ & $\cdot 10^{-5}$ & $\cdot 10^{-5}$ & &  \\
\hline
40 - 60 & 289 $\pm$ 2 & 297 $\pm$ 15 $\pm$ 19 & 5 & 6 \\
60 - 80 & 145 $\pm$ 2 & 128 $\pm$ 8 $\pm$ 13 & 6 & 10 \\
80 - 100 & 71 $\pm$ 1 & 78 $\pm$ 8 $\pm$ 11 & 10 & 14 \\
100 - 120 & 34.7 $\pm$ 0.8 & 41 $\pm$ 5 $\pm$ 6 & 13 & 14 \\
120 - 140 & 18.9 $\pm$ 0.6 & 21 $\pm$ 4 $\pm$ 4 & 21 & 19 \\
140 - 180 & 7.4 $\pm$ 0.3 & 6 $\pm$ 1 $\pm$ 1 & 21 & 21 \\
180 - 250 & 1.54 $\pm$ 0.09 & 1.1 $\pm$ 0.4 $\pm$ 0.5 & 32 & 45 \\
250 - 350 & 0.17 $\pm$ 0.03 & - & - & - \\
\hline
 60$\degree$ & $\cdot 10^{-5}$   & $\cdot 10^{-5}$  & & \\
\hline
40 - 60 & 1492 $\pm$ 6 & 1127 $\pm$ 24 $\pm$ 59 & 2 & 5 \\
60 - 80 & 1111 $\pm$ 5 & 1100 $\pm$ 26 $\pm$ 57 & 2 & 5 \\
80 - 100 & 766 $\pm$ 4 & 808 $\pm$ 23 $\pm$ 37 & 3 & 5 \\
100 - 120 & 472 $\pm$ 3 & 534 $\pm$ 20 $\pm$ 24 & 4 & 5 \\
120 - 140 & 252 $\pm$ 2 & 371 $\pm$ 18 $\pm$ 16 & 5 & 4 \\
140 - 160 & 129 $\pm$ 2 & 262 $\pm$ 18 $\pm$ 22 & 7 & 9 \\
160 - 180 & 64 $\pm$ 1 & 189 $\pm$ 19 $\pm$ 35 & 10 & 18 \\
180 - 200 & 35.2 $\pm$ 0.9 & 159 $\pm$ 25 $\pm$ 22 & 16 & 14 \\
200 - 230 & 18.4 $\pm$ 0.5 & 125 $\pm$ 24 $\pm$ 42 & 19 & 33 \\
230 - 260 & 8.7 $\pm$ 0.4 & 69 $\pm$ 20 $\pm$ 30 & 29 & 44 \\
260 - 290 & 3.9 $\pm$ 0.2 & 29 $\pm$ 11 $\pm$ 8 & 38 & 27 \\
290 - 350 & 1.4 $\pm$ 0.1 & 6 $\pm$ 2 $\pm$ 3 & 31 & 52 \\
350 - 450 & 0.32 $\pm$ 0.04 & - & - & - \\
\hline
\end{tabular}
\end{center}
\caption{Double differential cross section in kinetic energy bins at production ($E^p_{kin}$) of protons produced by 221~MeV/u $^{12}$C ion beam impinging on a CH target, detected at 90$\degree$ (top panel) and 60$\degree$ (bottom panel). The production cross section from the FLUKA Monte Carlo simulation (\textit{MC}) is listed alongside the experimental cross section (\textit{data}), with the relative statistical and systematic data uncertainties reported as percentage in the last two columns. }
\end{minipage}
\begin{minipage}{\columnwidth}
\begin{center}
\begin{tabular}{|c||cccc|}
\hline
\rule[-4mm]{0mm}{1cm}
 $E^p_{kin}$  & $\frac{d^2\sigma^{MC}}{d\Omega dE_k}$ & $\frac{d^2\sigma^{data}}{d\Omega dE_k}$ & $\Delta_{stat}^{data}$ & $\Delta_{sys}^{data}$\\
 $[$MeV$]$ & [b/sr/MeV] &  [b/sr/MeV] & [$\%$] & [$\%$]\\
\hline
 90$\degree$ & $\cdot 10^{-5}$  & $\cdot 10^{-5}$  & &  \\
\hline
40 - 60 & 351 $\pm$ 2 & 330 $\pm$ 14 $\pm$ 14 & 4 & 4 \\
60 - 80 & 179 $\pm$ 2 & 208 $\pm$ 12 $\pm$ 9 & 6 & 4 \\
80 - 100 & 93 $\pm$ 1 & 115 $\pm$ 9 $\pm$ 7 & 8 & 6 \\
100 - 120 & 48.4 $\pm$ 0.9 & 68 $\pm$ 7 $\pm$ 8 & 11 & 12 \\
120 - 140 & 23.8 $\pm$ 0.6 & 27 $\pm$ 4 $\pm$ 6 & 15 & 22 \\
140 - 180 & 9.4 $\pm$ 0.3 & 18 $\pm$ 4 $\pm$ 3 & 20 & 14 \\
180 - 250 & 1.64 $\pm$ 0.09 & 3.3 $\pm$ 1.3 $\pm$ 0.7 & 38 & 20 \\
250 - 350 & 0.16 $\pm$ 0.02 & - & - & - \\
\hline
 60$\degree$ & $\cdot 10^{-5}$ & $\cdot 10^{-5}$ & &\\
\hline
40 - 60 & 1662 $\pm$ 6 & 1142 $\pm$ 23 $\pm$ 63 & 2 & 5 \\
60 - 80 & 1333 $\pm$ 5 & 1195 $\pm$ 25 $\pm$ 69 & 2 & 6 \\
80 - 100 & 1042 $\pm$ 5 & 938 $\pm$ 23 $\pm$ 50 & 2 & 5 \\
100 - 120 & 744 $\pm$ 4 & 686 $\pm$ 20 $\pm$ 32 & 3 & 5 \\
120 - 140 & 484 $\pm$ 3 & 513 $\pm$ 18 $\pm$ 43 & 3 & 8 \\
140 - 160 & 286 $\pm$ 2 & 387 $\pm$ 18 $\pm$ 37 & 5 & 9 \\
160 - 180 & 169 $\pm$ 2 & 303 $\pm$ 19 $\pm$ 24 & 6 & 8 \\
180 - 200 & 92 $\pm$ 1 & 293 $\pm$ 29 $\pm$ 92 & 10 & 31 \\
200 - 230 & 46.4 $\pm$ 0.8 & 222 $\pm$ 26 $\pm$ 100 & 12 & 45 \\
230 - 260 & 20.1 $\pm$ 0.5 & 118 $\pm$ 20 $\pm$ 45 & 17 & 38 \\
260 - 290 & 8.9 $\pm$ 0.3 & 72 $\pm$ 20 $\pm$ 27 & 28 & 38 \\
290 - 350 & 3.0 $\pm$ 0.1 & 16 $\pm$ 4 $\pm$ 6 & 24 & 39 \\
350 - 450 & 0.37 $\pm$ 0.04 & 4 $\pm$ 3 $\pm$ 3 & 67 & 64 \\
\hline
\end{tabular}
\end{center}
\caption{Double differential cross section in kinetic energy bins at production ($E^p_{kin}$) of protons produced by 279~MeV/u $^{12}$C ion beam impinging on a CH target, detected at 90$\degree$ (top panel) and 60$\degree$ (bottom panel). The production cross section from the FLUKA Monte Carlo simulation (\textit{MC}) is listed alongside the experimental cross section (\textit{data}), with the relative statistical and systematic data uncertainties reported as percentage in the last two columns. }
\end{minipage}
\end{table*}

\begin{table*}
\begin{minipage}{\columnwidth}
\begin{center}
\begin{tabular}{|c||cccc|}
\hline
\rule[-4mm]{0mm}{1cm}
 $E^p_{kin}$  & $\frac{d^2\sigma^{MC}}{d\Omega dE_k}$ & $\frac{d^2\sigma^{data}}{d\Omega dE_k}$ & $\Delta_{stat}^{data}$ & $\Delta_{sys}^{data}$\\
 $[$MeV$]$ & [b/sr/MeV] &  [b/sr/MeV] & [$\%$] & [$\%$]\\
\hline
 90$\degree$ & $\cdot 10^{-5}$ & $\cdot 10^{-5}$  & & \\
\hline
40 - 60 & 391 $\pm$ 3 & 355 $\pm$ 14 $\pm$ 21 & 4 & 6 \\
60 - 80 & 213 $\pm$ 2 & 222 $\pm$ 12 $\pm$ 15 & 5 & 7 \\
80 - 100 & 124 $\pm$ 1 & 145 $\pm$ 10 $\pm$ 8 & 7 & 6 \\
100 - 120 & 71 $\pm$ 1 & 78 $\pm$ 7 $\pm$ 14 & 9 & 18 \\
120 - 140 & 38.2 $\pm$ 0.8 & 56 $\pm$ 8 $\pm$ 6 & 14 & 10 \\
140 - 180 & 15.6 $\pm$ 0.4 & 21 $\pm$ 3 $\pm$ 2 & 13 & 8 \\
180 - 250 & 3.1 $\pm$ 0.1 & 8 $\pm$ 2 $\pm$ 1 & 28 & 14 \\
250 - 350 & 0.31 $\pm$ 0.03 & 0.39 $\pm$ 0.17 $\pm$ 0.08 & 44 & 21 \\
\hline
 60$\degree$ & $\cdot 10^{-5}$  & $\cdot 10^{-5}$  & &  \\
\hline
40 - 60 & 1603 $\pm$ 6 & 1030 $\pm$ 20 $\pm$ 76 & 2 & 7 \\
60 - 80 & 1395 $\pm$ 5 & 1269 $\pm$ 27 $\pm$ 81 & 2 & 6 \\
80 - 100 & 1190 $\pm$ 5 & 986 $\pm$ 22 $\pm$ 51 & 2 & 5 \\
100 - 120 & 955 $\pm$ 4 & 693 $\pm$ 17 $\pm$ 38 & 2 & 5 \\
120 - 140 & 716 $\pm$ 4 & 612 $\pm$ 18 $\pm$ 44 & 3 & 7 \\
140 - 160 & 487 $\pm$ 3 & 526 $\pm$ 19 $\pm$ 28 & 4 & 5 \\
160 - 180 & 319 $\pm$ 3 & 403 $\pm$ 18 $\pm$ 47 & 5 & 12 \\
180 - 200 & 203 $\pm$ 2 & 324 $\pm$ 20 $\pm$ 51 & 6 & 16 \\
200 - 230 & 112 $\pm$ 1 & 340 $\pm$ 26 $\pm$ 163 & 8 & 48 \\
230 - 260 & 55.8 $\pm$ 0.9 & 211 $\pm$ 23 $\pm$ 91 & 11 & 43 \\
260 - 290 & 28.3 $\pm$ 0.6 & 82 $\pm$ 10 $\pm$ 19 & 12 & 23 \\
290 - 350 & 10.3 $\pm$ 0.3 & 47 $\pm$ 7 $\pm$ 11 & 14 & 24 \\
350 - 450 & 1.95 $\pm$ 0.09 & 29 $\pm$ 13 $\pm$ 5 & 44 & 16 \\
450 - 650 & 0.08 $\pm$ 0.01 & - & - & - \\
\hline
\end{tabular}
\end{center}
\caption{Double differential cross section in kinetic energy bins at production ($E^p_{kin}$) of protons produced by 351~MeV/u $^{12}$C ion beam impinging on a CH target, detected at 90$\degree$ (top panel) and 60$\degree$ (bottom panel). The production cross section from the FLUKA Monte Carlo simulation (\textit{MC}) is listed alongside the experimental cross section (\textit{data}), with the relative statistical and systematic data uncertainties reported as percentage in the last two columns. }
\end{minipage}
\begin{minipage}{\columnwidth}
\begin{center}
\begin{tabular}{|c||cccc|}
\hline
\rule[-4mm]{0mm}{1cm}
 $E^C_{kin}$ & $\frac{d\sigma^{MC}}{d\Omega}$ &  $\frac{d\sigma^{data}}{d\Omega}$ & $\Delta_{stat}^{data}$ & $\Delta_{sys}^{data}$ \\
 $[$MeV/u$]$  & [b/sr] & [b/sr] & [$\%$] &  [$\%$]\\
\hline
 90$\degree$ & $\cdot 10^{-3}$ & $\cdot 10^{-3}$ & & \\
\hline
115 & 55.0 $\pm$ 0.4 & 42 $\pm$ 2 $\pm$ 2 & 5 & 4 \\
150 & 78.2 $\pm$ 0.5 & 62 $\pm$ 2 $\pm$ 2 & 4 & 4 \\
221 & 115.8 $\pm$ 0.7 & 115 $\pm$ 4 $\pm$ 5 & 3 & 4 \\
279 & 144.0 $\pm$ 0.7 & 157 $\pm$ 5 $\pm$ 6 & 3 & 4 \\
351 & 175.7 $\pm$ 0.8 & 183 $\pm$ 5 $\pm$ 8 & 3 & 4 \\
\hline
 60$\degree$ & $\cdot 10^{-2}$ & $\cdot 10^{-2}$ &  & \\
\hline
115 & 42.1 $\pm$ 0.1 & 43.2 $\pm$ 0.7 $\pm$ 2.1 & 2 & 5 \\
150 & 44.9 $\pm$ 0.1 & 64 $\pm$ 1 $\pm$ 3 & 2 & 4 \\
221 & 87.4 $\pm$ 0.2 & 94 $\pm$ 1 $\pm$ 4 & 1 & 5 \\
279 & 118.7 $\pm$ 0.2 & 115 $\pm$ 1 $\pm$ 5 & 1 & 5 \\
351 & 143.9 $\pm$ 0.2 & 130 $\pm$ 1 $\pm$ 7 & 1 & 6 \\
\hline
\end{tabular}
\end{center}
\caption{Energy integrated differential cross section of protons produced by $^{12}$C ion beam of kinetic energy ($E^C_{kin}$) ranging from 115~MeV/u to 351~MeV/u impinging on a CH target. Protons are detected at 90$\degree$ (top panel) and 60$\degree$ (bottom panel). The production cross section from the FLUKA Monte Carlo simulation (\textit{MC}) is listed alongside the experimental cross section (\textit{data}), with the relative statistical and systematic data uncertainties reported as percentage in the last two columns.}
\end{minipage}
\end{table*}

\begin{table*}
\begin{minipage}{\columnwidth}
\begin{center}
\begin{tabular}{|c||cccc|}
\hline
\rule[-4mm]{0mm}{1cm}
 $E^C_{kin}$ & $\frac{d\sigma^{MC}}{d\Omega}$ &  $\frac{d\sigma^{data}}{d\Omega}$ & $\Delta_{stat}^{data}$ & $\Delta_{sys}^{data}$ \\
 $[$MeV/u$]$  & [b/sr] & [b/sr] & [$\%$] &  [$\%$]\\
\hline
 90$\degree$ & $\cdot 10^{-4}$ & $\cdot 10^{-4}$ & & \\
\hline
150 & 86 $\pm$ 2 & 29 $\pm$ 4 $\pm$ 3 & 15 & 11 \\
221 & 144 $\pm$ 2 & 58 $\pm$ 6 $\pm$ 5 & 11 & 8 \\
279 & 130 $\pm$ 2 & 79 $\pm$ 8 $\pm$ 7 & 10 & 9 \\
351 & 112 $\pm$ 2 & 115 $\pm$ 12 $\pm$ 14 & 10 & 12 \\
\hline
 60$\degree$ & $\cdot 10^{-3}$ & $\cdot 10^{-3}$ & & \\
\hline
115 & 56.0 $\pm$ 0.5 & 37 $\pm$ 2 $\pm$ 3 & 5 & 7 \\
150 & 61.7 $\pm$ 0.5 & 66 $\pm$ 3 $\pm$ 4 & 4 & 6 \\
221 & 63.2 $\pm$ 0.5 & 106 $\pm$ 5 $\pm$ 6 & 4 & 5 \\
279 & 48.6 $\pm$ 0.4 & 132 $\pm$ 7 $\pm$ 9 & 5 & 6 \\
351 & 39.2 $\pm$ 0.4 & 116 $\pm$ 6 $\pm$ 11 & 5 & 9 \\
\hline
\end{tabular}
\end{center}
\caption{Energy integrated differential cross section of deuterons produced by $^{12}$C ion beam of kinetic energy ($E^C_{kin}$) ranging from 115~MeV/u to 351~MeV/u impinging on a CH target. Deuterons are detected at 90$\degree$ (top panel) and 60$\degree$ (bottom panel). The production cross section from the FLUKA Monte Carlo simulation (\textit{MC}) is listed alongside the experimental cross section (\textit{data}), with the relative statistical and systematic data uncertainties reported as percentage in the last two columns. Cross section of 115~MeV/u $^{12}$C ion beam impinging on a CH target for deuterons detected at 90$\degree$ is not reported due to insufficient statistics.}
\end{minipage}
\begin{minipage}{\columnwidth}
\begin{center}
\begin{tabular}{|c||cccc|}
\hline
\rule[-4mm]{0mm}{1cm}
 $E^C_{kin}$ & $\frac{d\sigma^{MC}}{d\Omega}$ &  $\frac{d\sigma^{data}}{d\Omega}$ & $\Delta_{stat}^{data}$ & $\Delta_{sys}^{data}$ \\
 $[$MeV/u$]$  & [b/sr] & [b/sr] & [$\%$] &  [$\%$]\\
\hline
 60$\degree$ & $\cdot 10^{-4}$  & $\cdot 10^{-4}$  & & \\
\hline
115 & 44 $\pm$ 1 & 31 $\pm$ 5 $\pm$ 8 & 17 & 25 \\
150 & 56 $\pm$ 1 & 67 $\pm$ 10 $\pm$ 5 & 15 & 7 \\
221 & 75 $\pm$ 2 & 127 $\pm$ 17 $\pm$ 15 & 14 & 12 \\
279 & 58 $\pm$ 2 & 172 $\pm$ 24 $\pm$ 11 & 14 & 6 \\
351 & 42 $\pm$ 1 & 185 $\pm$ 29 $\pm$ 42 & 16 & 23 \\
\hline
\end{tabular}
\end{center}
\caption{Energy integrated differential cross section of tritons produced by $^{12}$C ion beam of kinetic energy ($E^C_{kin}$) ranging from 115~MeV/u to 351~MeV/u impinging on a CH target. Tritons are detected at 60$\degree$. The production cross section from the FLUKA Monte Carlo simulation (\textit{MC}) is listed alongside the experimental cross section (\textit{data}), with the relative statistical and systematic data uncertainties reported as percentage in the last two columns. Cross section for tritons detected at 90$\degree$ is not reported due to insufficient statistics.}
\end{minipage}
\end{table*}

\begin{table*}
\begin{minipage}{\columnwidth}
\begin{center}
\begin{tabular}{|c||cccc|}
\hline
\rule[-4mm]{0mm}{1cm}
$E^p_{kin}$  & $\frac{d^2\sigma^{MC}}{d\Omega dE_k}$ & $\frac{d^2\sigma^{data}}{d\Omega dE_k}$ & $\Delta_{stat}^{data}$ & $\Delta_{sys}^{data}$\\
 $[$MeV$]$ & [b/sr/MeV] &  [b/sr/MeV] & [$\%$] & [$\%$]\\
\hline
 90$\degree$ & $\cdot 10^{-4}$  & $\cdot 10^{-4}$  & & \\
\hline
40 - 60 & 13.4 $\pm$ 0.1 & 12.0 $\pm$ 0.7 $\pm$ 0.7 & 5 & 6 \\
60 - 80 & 5.76 $\pm$ 0.09 & 4.5 $\pm$ 0.4 $\pm$ 0.3 & 9 & 7 \\
80 - 100 & 3.01 $\pm$ 0.06 & 1.48 $\pm$ 0.20 $\pm$ 0.06 & 13 & 4 \\
100 - 120 & 1.48 $\pm$ 0.04 & 0.46 $\pm$ 0.08 $\pm$ 0.08 & 18 & 17 \\
120 - 140 & 0.77 $\pm$ 0.03 & - & - & - \\
140 - 180 & 0.26 $\pm$ 0.01 & - & - & - \\
180 - 250 & 0.046 $\pm$ 0.004 & - & - & - \\
\hline
 60$\degree$ & $\cdot 10^{-4}$   & $\cdot 10^{-4}$ & & \\
\hline
40 - 60 & 82.0 $\pm$ 0.3 & 79 $\pm$ 2 $\pm$ 4 & 2 & 5 \\
60 - 80 & 49.8 $\pm$ 0.3 & 57 $\pm$ 2 $\pm$ 2 & 3 & 4 \\
80 - 100 & 24.7 $\pm$ 0.2 & 29 $\pm$ 1 $\pm$ 2 & 4 & 6 \\
100 - 120 & 11.2 $\pm$ 0.1 & 12.9 $\pm$ 0.8 $\pm$ 0.6 & 6 & 5 \\
120 - 140 & 5.69 $\pm$ 0.09 & 8.4 $\pm$ 0.8 $\pm$ 0.5 & 9 & 5 \\
140 - 160 & 2.97 $\pm$ 0.07 & 5.8 $\pm$ 0.8 $\pm$ 1.2 & 14 & 21 \\
160 - 180 & 1.78 $\pm$ 0.05 & 2.0 $\pm$ 0.3 $\pm$ 0.6 & 15 & 31 \\
180 - 200 & 0.95 $\pm$ 0.04 & 1.9 $\pm$ 0.5 $\pm$ 0.7 & 28 & 36 \\
200 - 230 & 0.48 $\pm$ 0.02 & 0.8 $\pm$ 0.2 $\pm$ 0.2 & 29 & 22 \\
230 - 260 & 0.19 $\pm$ 0.01 & 1.7 $\pm$ 1.6 $\pm$ 0.3 & 95 & 21 \\
260 - 290 & 0.078 $\pm$ 0.009 & - & - & - \\
290 - 350 & 0.026 $\pm$ 0.004 & - & - & - \\
\hline
\end{tabular}
\end{center}
\caption{Double differential cross section in kinetic energy bins at production ($E^p_{kin}$) of protons produced by 115~MeV/u $^{12}$C ion beam impinging on a PMMA target, detected at 90$\degree$ (top panel) and 60$\degree$ (bottom panel). The production cross section from the FLUKA Monte Carlo simulation (\textit{MC}) is listed alongside the experimental cross section (\textit{data}), with the relative statistical and systematic data uncertainties reported as percentage in the last two columns. }
\end{minipage}
\begin{minipage}{\columnwidth}
\begin{center}
\begin{tabular}{|c||cccc|}
\hline
\rule[-4mm]{0mm}{1cm}
$E^p_{kin}$  & $\frac{d^2\sigma^{MC}}{d\Omega dE_k}$ & $\frac{d^2\sigma^{data}}{d\Omega dE_k}$ & $\Delta_{stat}^{data}$ & $\Delta_{sys}^{data}$\\
 $[$MeV$]$ & [b/sr/MeV] &  [b/sr/MeV] & [$\%$] & [$\%$]\\
\hline
 90$\degree$ & $\cdot 10^{-4}$  & $\cdot 10^{-4}$ & &  \\
\hline
40 - 60 & 17.3 $\pm$ 0.1 & 17.0 $\pm$ 0.8 $\pm$ 1.2 & 5 & 7 \\
60 - 80 & 7.9 $\pm$ 0.1 & 7.9 $\pm$ 0.6 $\pm$ 0.7 & 8 & 9 \\
80 - 100 & 3.77 $\pm$ 0.07 & 2.8 $\pm$ 0.3 $\pm$ 0.1 & 10 & 5 \\
100 - 120 & 1.82 $\pm$ 0.05 & 1.5 $\pm$ 0.3 $\pm$ 0.4 & 17 & 24 \\
120 - 140 & 0.84 $\pm$ 0.03 & 0.41 $\pm$ 0.09 $\pm$ 0.10 & 22 & 24 \\
140 - 180 & 0.32 $\pm$ 0.01 & 0.13 $\pm$ 0.03 $\pm$ 0.03 & 27 & 27 \\
180 - 250 & 0.059 $\pm$ 0.005 & - & - & - \\
\hline
 60$\degree$ & $\cdot 10^{-4}$  & $\cdot 10^{-4}$  & & \\
\hline
40 - 60 & 77.2 $\pm$ 0.3 & 83 $\pm$ 2 $\pm$ 4 & 2 & 5 \\
60 - 80 & 51.8 $\pm$ 0.3 & 76 $\pm$ 2 $\pm$ 3 & 3 & 4 \\
80 - 100 & 28.8 $\pm$ 0.2 & 48 $\pm$ 2 $\pm$ 2 & 4 & 4 \\
100 - 120 & 13.6 $\pm$ 0.1 & 26 $\pm$ 1 $\pm$ 1 & 5 & 4 \\
120 - 140 & 6.63 $\pm$ 0.10 & 17 $\pm$ 1 $\pm$ 1 & 8 & 8 \\
140 - 160 & 3.70 $\pm$ 0.07 & 9.5 $\pm$ 1.0 $\pm$ 0.9 & 11 & 10 \\
160 - 180 & 2.16 $\pm$ 0.06 & 6.3 $\pm$ 0.9 $\pm$ 0.8 & 15 & 13 \\
180 - 200 & 1.28 $\pm$ 0.04 & 6.0 $\pm$ 1.5 $\pm$ 0.8 & 25 & 14 \\
200 - 230 & 0.66 $\pm$ 0.03 & 2.8 $\pm$ 0.8 $\pm$ 1.5 & 27 & 55 \\
230 - 260 & 0.34 $\pm$ 0.02 & 1.32 $\pm$ 0.47 $\pm$ 0.06 & 35 & 5 \\
260 - 290 & 0.15 $\pm$ 0.01 & 0.33 $\pm$ 0.14 $\pm$ 0.03 & 41 & 10 \\
290 - 350 & 0.044 $\pm$ 0.005 & - & - & - \\
350 - 450 & 0.007 $\pm$ 0.001 & - & - & - \\
\hline
\end{tabular}
\end{center}
\caption{Double differential cross section in kinetic energy bins at production ($E^p_{kin}$) of protons produced by 150~MeV/u $^{12}$C ion beam impinging on a PMMA target, detected at 90$\degree$ (top panel) and 60$\degree$ (bottom panel). The production cross section from the FLUKA Monte Carlo simulation (\textit{MC}) is listed alongside the experimental cross section (\textit{data}), with the relative statistical and systematic data uncertainties reported as percentage in the last two columns. }
\end{minipage}
\end{table*}

\begin{table*}
\begin{minipage}{\columnwidth}
\begin{center}
\begin{tabular}{|c||cccc|}
\hline
\rule[-4mm]{0mm}{1cm}
$E^p_{kin}$  & $\frac{d^2\sigma^{MC}}{d\Omega dE_k}$ & $\frac{d^2\sigma^{data}}{d\Omega dE_k}$ & $\Delta_{stat}^{data}$ & $\Delta_{sys}^{data}$\\
 $[$MeV$]$ & [b/sr/MeV] &  [b/sr/MeV] & [$\%$] & [$\%$]\\
\hline
 90$\degree$ & $\cdot 10^{-4}$  & $\cdot 10^{-4}$ & &  \\
\hline
40 - 60 & 24.2 $\pm$ 0.2 & 23.8 $\pm$ 1.0 $\pm$ 1.0 & 4 & 4 \\
60 - 80 & 12.0 $\pm$ 0.1 & 12.9 $\pm$ 0.7 $\pm$ 0.9 & 6 & 7 \\
80 - 100 & 5.88 $\pm$ 0.09 & 6.9 $\pm$ 0.6 $\pm$ 0.4 & 9 & 6 \\
100 - 120 & 2.92 $\pm$ 0.06 & 3.3 $\pm$ 0.4 $\pm$ 0.3 & 11 & 9 \\
120 - 140 & 1.64 $\pm$ 0.05 & 1.7 $\pm$ 0.3 $\pm$ 0.2 & 16 & 13 \\
140 - 180 & 0.66 $\pm$ 0.02 & 0.8 $\pm$ 0.2 $\pm$ 0.2 & 21 & 30 \\
180 - 250 & 0.147 $\pm$ 0.007 & 0.09 $\pm$ 0.03 $\pm$ 0.03 & 32 & 28 \\
250 - 350 & 0.013 $\pm$ 0.002 & - & - & - \\
\hline
 60$\degree$ & $\cdot 10^{-4}$  & $\cdot 10^{-4}$  & & \\
\hline
40 - 60 & 124.7 $\pm$ 0.4 & 101 $\pm$ 2 $\pm$ 6 & 2 & 6 \\
60 - 80 & 92.8 $\pm$ 0.4 & 97 $\pm$ 2 $\pm$ 6 & 2 & 6 \\
80 - 100 & 64.5 $\pm$ 0.3 & 69 $\pm$ 2 $\pm$ 3 & 3 & 5 \\
100 - 120 & 39.2 $\pm$ 0.2 & 47 $\pm$ 2 $\pm$ 2 & 3 & 5 \\
120 - 140 & 21.4 $\pm$ 0.2 & 37 $\pm$ 2 $\pm$ 2 & 5 & 5 \\
140 - 160 & 10.8 $\pm$ 0.1 & 25 $\pm$ 2 $\pm$ 3 & 7 & 11 \\
160 - 180 & 5.49 $\pm$ 0.09 & 22 $\pm$ 2 $\pm$ 4 & 10 & 20 \\
180 - 200 & 3.06 $\pm$ 0.07 & 13 $\pm$ 2 $\pm$ 5 & 13 & 37 \\
200 - 230 & 1.51 $\pm$ 0.04 & 9 $\pm$ 2 $\pm$ 3 & 17 & 35 \\
230 - 260 & 0.70 $\pm$ 0.03 & 15 $\pm$ 6 $\pm$ 5 & 40 & 34 \\
260 - 290 & 0.35 $\pm$ 0.02 & 3.2 $\pm$ 1.1 $\pm$ 0.7 & 35 & 23 \\
290 - 350 & 0.137 $\pm$ 0.008 & 0.7 $\pm$ 0.2 $\pm$ 0.2 & 29 & 26 \\
350 - 450 & 0.027 $\pm$ 0.003 & 0.3 $\pm$ 0.2 $\pm$ 0.2 & 69 & 61 \\
\hline
\end{tabular}
\end{center}
\caption{Double differential cross section in kinetic energy bins at production ($E^p_{kin}$) of protons produced by 221~MeV/u $^{12}$C ion beam impinging on a PMMA target, detected at 90$\degree$ (top panel) and 60$\degree$ (bottom panel). The production cross section from the FLUKA Monte Carlo simulation (\textit{MC}) is listed alongside the experimental cross section (\textit{data}), with the relative statistical and systematic data uncertainties reported as percentage in the last two columns. }
\end{minipage}
\begin{minipage}{\columnwidth}
\begin{center}
\begin{tabular}{|c||cccc|}
\hline
\rule[-4mm]{0mm}{1cm}
$E^p_{kin}$  & $\frac{d^2\sigma^{MC}}{d\Omega dE_k}$ & $\frac{d^2\sigma^{data}}{d\Omega dE_k}$ & $\Delta_{stat}^{data}$ & $\Delta_{sys}^{data}$\\
 $[$MeV$]$ & [b/sr/MeV] &  [b/sr/MeV] & [$\%$] & [$\%$]\\
\hline
 90$\degree$ & $\cdot 10^{-4}$ & $\cdot 10^{-4}$  & &  \\
\hline
40 - 60 & 29.9 $\pm$ 0.2 & 28 $\pm$ 1 $\pm$ 1 & 4 & 5 \\
60 - 80 & 15.1 $\pm$ 0.2 & 16.4 $\pm$ 0.9 $\pm$ 0.7 & 6 & 4 \\
80 - 100 & 7.7 $\pm$ 0.1 & 9.4 $\pm$ 0.8 $\pm$ 0.5 & 8 & 6 \\
100 - 120 & 4.28 $\pm$ 0.08 & 4.8 $\pm$ 0.5 $\pm$ 0.2 & 10 & 5 \\
120 - 140 & 2.15 $\pm$ 0.06 & 3.4 $\pm$ 0.6 $\pm$ 0.9 & 17 & 25 \\
140 - 180 & 0.83 $\pm$ 0.03 & 1.3 $\pm$ 0.3 $\pm$ 0.2 & 19 & 12 \\
180 - 250 & 0.164 $\pm$ 0.009 & 0.30 $\pm$ 0.09 $\pm$ 0.04 & 31 & 12 \\
250 - 350 & 0.015 $\pm$ 0.002 & - & - & - \\
\hline
 60$\degree$ & $\cdot 10^{-4}$  & $\cdot 10^{-4}$  & & \\
\hline
40 - 60 & 140.5 $\pm$ 0.5 & 101 $\pm$ 2 $\pm$ 7 & 2 & 7 \\
60 - 80 & 112.2 $\pm$ 0.5 & 105 $\pm$ 2 $\pm$ 7 & 2 & 7 \\
80 - 100 & 87.4 $\pm$ 0.4 & 80 $\pm$ 2 $\pm$ 4 & 2 & 6 \\
100 - 120 & 62.5 $\pm$ 0.3 & 60 $\pm$ 2 $\pm$ 3 & 3 & 5 \\
120 - 140 & 40.3 $\pm$ 0.3 & 44 $\pm$ 2 $\pm$ 3 & 4 & 8 \\
140 - 160 & 24.0 $\pm$ 0.2 & 36 $\pm$ 2 $\pm$ 2 & 5 & 5 \\
160 - 180 & 13.9 $\pm$ 0.2 & 31 $\pm$ 2 $\pm$ 3 & 7 & 10 \\
180 - 200 & 7.7 $\pm$ 0.1 & 22 $\pm$ 2 $\pm$ 3 & 10 & 12 \\
200 - 230 & 3.83 $\pm$ 0.07 & 17 $\pm$ 2 $\pm$ 7 & 12 & 41 \\
230 - 260 & 1.70 $\pm$ 0.05 & 10 $\pm$ 2 $\pm$ 5 & 17 & 46 \\
260 - 290 & 0.78 $\pm$ 0.03 & 4.5 $\pm$ 1.0 $\pm$ 1.6 & 22 & 35 \\
290 - 350 & 0.26 $\pm$ 0.01 & 1.8 $\pm$ 0.5 $\pm$ 0.7 & 27 & 40 \\
350 - 450 & 0.040 $\pm$ 0.004 & - & - & - \\
\hline
\end{tabular}
\end{center}
\caption{Double differential cross section in kinetic energy bins at production ($E^p_{kin}$) of protons produced by 279~MeV/u $^{12}$C ion beam impinging on a PMMA target, detected at 90$\degree$ (top panel) and 60$\degree$ (bottom panel). The production cross section from the FLUKA Monte Carlo simulation (\textit{MC}) is listed alongside the experimental cross section (\textit{data}), with the relative statistical and systematic data uncertainties reported as percentage in the last two columns. }
\end{minipage}
\end{table*}

\begin{table*}
\begin{minipage}{\columnwidth}
\begin{center}
\begin{tabular}{|c||cccc|}
\hline
\rule[-4mm]{0mm}{1cm}
$E^p_{kin}$  & $\frac{d^2\sigma^{MC}}{d\Omega dE_k}$ & $\frac{d^2\sigma^{data}}{d\Omega dE_k}$ & $\Delta_{stat}^{data}$ & $\Delta_{sys}^{data}$\\
 $[$MeV$]$ & [b/sr/MeV] &  [b/sr/MeV] & [$\%$] & [$\%$]\\
\hline
 90$\degree$ & $\cdot 10^{-4}$ & $\cdot 10^{-4}$ & & \\
\hline
40 - 60 & 33.8 $\pm$ 0.2 & 32 $\pm$ 1 $\pm$ 1 & 4 & 4 \\
60 - 80 & 19.0 $\pm$ 0.2 & 19.9 $\pm$ 1.0 $\pm$ 1.1 & 5 & 6 \\
80 - 100 & 10.4 $\pm$ 0.1 & 11.2 $\pm$ 0.7 $\pm$ 0.7 & 6 & 6 \\
100 - 120 & 6.06 $\pm$ 0.09 & 8.2 $\pm$ 0.7 $\pm$ 0.4 & 9 & 5 \\
120 - 140 & 3.43 $\pm$ 0.07 & 5.0 $\pm$ 0.6 $\pm$ 0.2 & 12 & 4 \\
140 - 180 & 1.43 $\pm$ 0.03 & 2.0 $\pm$ 0.2 $\pm$ 0.3 & 12 & 14 \\
180 - 250 & 0.30 $\pm$ 0.01 & 0.6 $\pm$ 0.1 $\pm$ 0.2 & 22 & 24 \\
250 - 350 & 0.026 $\pm$ 0.003 & - & - & - \\
\hline
 60$\degree$ & $\cdot 10^{-4}$  & $\cdot 10^{-4}$ &&  \\
\hline
40 - 60 & 136.1 $\pm$ 0.5 & 97 $\pm$ 2 $\pm$ 5 & 2 & 5 \\
60 - 80 & 117.8 $\pm$ 0.4 & 107 $\pm$ 2 $\pm$ 6 & 2 & 5 \\
80 - 100 & 100.8 $\pm$ 0.4 & 89 $\pm$ 2 $\pm$ 6 & 2 & 6 \\
100 - 120 & 80.7 $\pm$ 0.4 & 65 $\pm$ 2 $\pm$ 3 & 2 & 5 \\
120 - 140 & 60.6 $\pm$ 0.3 & 55 $\pm$ 2 $\pm$ 4 & 3 & 7 \\
140 - 160 & 41.1 $\pm$ 0.3 & 45 $\pm$ 2 $\pm$ 4 & 4 & 9 \\
160 - 180 & 26.8 $\pm$ 0.2 & 36 $\pm$ 2 $\pm$ 2 & 4 & 5 \\
180 - 200 & 17.1 $\pm$ 0.2 & 33 $\pm$ 2 $\pm$ 4 & 6 & 12 \\
200 - 230 & 9.3 $\pm$ 0.1 & 29 $\pm$ 2 $\pm$ 11 & 8 & 39 \\
230 - 260 & 4.70 $\pm$ 0.07 & 17 $\pm$ 2 $\pm$ 3 & 10 & 18 \\
260 - 290 & 2.42 $\pm$ 0.05 & 10 $\pm$ 1 $\pm$ 1 & 13 & 14 \\
290 - 350 & 0.91 $\pm$ 0.02 & 3.7 $\pm$ 0.5 $\pm$ 1.5 & 14 & 40 \\
350 - 450 & 0.165 $\pm$ 0.007 & 2.3 $\pm$ 1.0 $\pm$ 1.0 & 44 & 43 \\
450 - 650 & 0.007 $\pm$ 0.001 & - & - & - \\
\hline
\end{tabular}
\end{center}
\caption{Double differential cross section in kinetic energy bins at production ($E^p_{kin}$) of protons produced by 351~MeV/u $^{12}$C ion beam impinging on a PMMA target, detected at 90$\degree$ (top panel) and 60$\degree$ (bottom panel). The production cross section from the FLUKA Monte Carlo simulation (\textit{MC}) is listed alongside the experimental cross section (\textit{data}), with the relative statistical and systematic data uncertainties reported as percentage in the last two columns. }
\end{minipage}
\begin{minipage}{\columnwidth}
\begin{center}
\begin{tabular}{|c||cccc|}
\hline
\rule[-4mm]{0mm}{1cm}
 $E^C_{kin}$ & $\frac{d\sigma^{MC}}{d\Omega}$ &  $\frac{d\sigma^{data}}{d\Omega}$ & $\Delta_{stat}^{data}$ & $\Delta_{sys}^{data}$ \\
 $[$MeV/u$]$  & [b/sr] & [b/sr] & [$\%$] &  [$\%$]\\
\hline
 90$\degree$ & $\cdot 10^{-2}$ & $\cdot 10^{-2}$ & & \\
\hline
115 & 5.01 $\pm$ 0.04 & 3.8 $\pm$ 0.2 $\pm$ 0.2 & 4 & 4 \\
150 & 6.49 $\pm$ 0.04 & 6.0 $\pm$ 0.2 $\pm$ 0.3 & 4 & 4 \\
221 & 9.70 $\pm$ 0.05 & 10.1 $\pm$ 0.3 $\pm$ 0.4 & 3 & 4 \\
279 & 12.28 $\pm$ 0.06 & 12.9 $\pm$ 0.4 $\pm$ 0.5 & 3 & 4 \\
351 & 15.32 $\pm$ 0.07 & 16.4 $\pm$ 0.4 $\pm$ 0.7 & 2 & 4 \\
\hline
 60$\degree$ & $\cdot 10^{-1}$ & $\cdot 10^{-1} $ &  & \\
\hline
115 & 3.61 $\pm$ 0.01 & 3.92 $\pm$ 0.06 $\pm$ 0.16 & 2 & 4 \\
150 & 3.74 $\pm$ 0.01 & 5.40 $\pm$ 0.08 $\pm$ 0.23 & 1 & 4 \\
221 & 7.32 $\pm$ 0.01 & 8.32 $\pm$ 0.09 $\pm$ 0.43 & 1 & 5 \\
279 & 9.98 $\pm$ 0.02 & 10.0 $\pm$ 0.1 $\pm$ 0.5 & 1 & 5 \\
351 & 12.17 $\pm$ 0.02 & 11.7 $\pm$ 0.1 $\pm$ 0.6 & 1 & 5 \\
\hline
\end{tabular}
\end{center}
\caption{Energy integrated differential cross section of protons produced by $^{12}$C ion beam of kinetic energy ($E^C_{kin}$) ranging from 115~MeV/u to 351~MeV/u impinging on a PMMA target. Protons are detected at 90$\degree$ (top panel) and 60$\degree$ (bottom panel). The production cross section from the FLUKA Monte Carlo simulation (\textit{MC}) is listed alongside the experimental cross section (\textit{data}), with the relative statistical and systematic data uncertainties reported as percentage in the last two columns.}
\end{minipage}
\end{table*}

\begin{table*}
\begin{minipage}{\columnwidth}
\begin{center}
\begin{tabular}{|c||cccc|}
\hline
\rule[-4mm]{0mm}{1cm}
 $E^C_{kin}$ & $\frac{d\sigma^{MC}}{d\Omega}$ &  $\frac{d\sigma^{data}}{d\Omega}$ & $\Delta_{stat}^{data}$ & $\Delta_{sys}^{data}$ \\
 $[$MeV/u$]$  & [b/sr] & [b/sr] & [$\%$] &  [$\%$]\\
\hline
 90$\degree$ & $\cdot 10^{-3}$ & $\cdot 10^{-3}$ & & \\
\hline
115 & 1.92 $\pm$ 0.07 & 0.8 $\pm$ 0.2 $\pm$ 0.1 & 23 & 14 \\
150 & 7.2 $\pm$ 0.1 & 2.1 $\pm$ 0.3 $\pm$ 0.1 & 14 & 7 \\
221 & 13.2 $\pm$ 0.2 & 4.1 $\pm$ 0.4 $\pm$ 0.3 & 9 & 8 \\
279 & 12.1 $\pm$ 0.2 & 8.1 $\pm$ 0.8 $\pm$ 0.4 & 10 & 5 \\
351 & 10.9 $\pm$ 0.2 & 10.3 $\pm$ 1.0 $\pm$ 1.0 & 10 & 10 \\
\hline
 60$\degree$ & $\cdot 10^{-2}$ & $\cdot 10^{-2}$  &  & \\
\hline
115 & 4.75 $\pm$ 0.04 & 3.5 $\pm$ 0.2 $\pm$ 0.3 & 5 & 8 \\
150 & 4.88 $\pm$ 0.04 & 6.0 $\pm$ 0.3 $\pm$ 0.4 & 4 & 7 \\
221 & 5.10 $\pm$ 0.04 & 10.3 $\pm$ 0.4 $\pm$ 0.7 & 4 & 6 \\
279 & 4.12 $\pm$ 0.04 & 13.1 $\pm$ 0.7 $\pm$ 0.7 & 5 & 6 \\
351 & 3.41 $\pm$ 0.03 & 12.0 $\pm$ 0.6 $\pm$ 0.7 & 5 & 6 \\
\hline
\end{tabular}
\end{center}
\caption{Energy integrated differential cross section of deuterons produced by $^{12}$C ion beam of kinetic energy ($E^C_{kin}$) ranging from 115~MeV/u to 351~MeV/u impinging on a PMMA target. Deuterons are detected at 90$\degree$ (top panel) and 60$\degree$ (bottom panel). The production cross section from the FLUKA Monte Carlo simulation (\textit{MC}) is listed alongside the experimental cross section (\textit{data}), with the relative statistical and systematic data uncertainties reported as percentage in the last two columns.}
\end{minipage}
\begin{minipage}{\columnwidth}
\begin{center}
\begin{tabular}{|c||cccc|}
\hline
\rule[-4mm]{0mm}{1cm}
 $E^C_{kin}$ & $\frac{d\sigma^{MC}}{d\Omega}$ &  $\frac{d\sigma^{data}}{d\Omega}$ & $\Delta_{stat}^{data}$ & $\Delta_{sys}^{data}$ \\
 $[$MeV/u$]$  & [b/sr] & [b/sr] & [$\%$] &  [$\%$]\\
\hline
 60$\degree$ & $\cdot 10^{-3}$ & $\cdot 10^{-3}$ && ]\\
\hline
115 & 3.25 $\pm$ 0.10 & 2.9 $\pm$ 0.6 $\pm$ 0.7 & 19 & 23 \\
150 & 4.6 $\pm$ 0.1 & 5.9 $\pm$ 0.9 $\pm$ 0.8 & 15 & 14 \\
221 & 6.1 $\pm$ 0.1 & 13.6 $\pm$ 1.6 $\pm$ 0.9 & 12 & 6 \\
279 & 5.1 $\pm$ 0.1 & 22 $\pm$ 4 $\pm$ 5 & 17 & 20 \\
351 & 4.1 $\pm$ 0.1 & 15 $\pm$ 2 $\pm$ 4 & 16 & 23 \\
\hline
\end{tabular}
\end{center}
\caption{Energy integrated differential cross section of tritons produced by $^{12}$C ion beam of kinetic energy ($E^C_{kin}$) ranging from 115~MeV/u to 351~MeV/u impinging on a PMMA target. Tritons are detected at 60$\degree$. The production cross section from the FLUKA Monte Carlo simulation (\textit{MC}) is listed alongside the experimental cross section (\textit{data}), with the relative statistical and systematic data uncertainties reported as percentage in the last two columns. Cross section for tritons detected at 90$\degree$ is not reported due to insufficient statistics.}
\end{minipage}
\end{table*}

%%%%%%

%\nocite{}

\end{document}